# Charged Dirac stars

Maribel Hernández-Márquez* and Miguel Alcubierre†
*Instituto de Ciencias Nucleares, Universidad Nacional Autónoma de México, Circuito Exterior C.U., A.P. 70-543, México D.F. 04510, México*

In this work we solve the coupled Einstein-Dirac-Maxwell (EDM) system for static spherically symmetric configurations of two fermions in a singlet spinor state within the 3+1 formalism of general relativity. We find different families of stationary self-gravitating solutions for the Dirac field through a numerical shooting method for different values of the electric charge parameter $q$. Furthermore, we investigate the effect of the charge $q$ on the binding energy, mass, radius, and compactness of the solutions. We show that gravitationally bound configurations exist only for $q < m$, with $m$ the mass of the Dirac field, and that the mass–frequency relation exhibits the characteristic spiral structure previously found for bosonic fields of spin $s = 0$ and $s = 1$. We are able to show that some of these gravitationally bound configurations have a compactness comparable to that of neutron stars. With these results, we conclude that at least at the classical level, self-gravitating fields with different spins $s = 0, 1/2, 1$ share some common characteristics when they are coupled to gravity. As has been previously shown for the case of bosonic stars, we also find some super-critical solutions with $q$ slightly larger than $m$. Such super-critical solutions correspond to configurations such that in the Newtonian regime the Coulomb repulsion overcomes the gravitational attraction, and as such they would not be expected to exist. Nevertheless, even if they do exist in the general relativistic case for a limited range of values of $q > m$, we find that they are always gravitationally unbound.

PACS numbers: 04.20.Ex, 04.25.Dm, 95.30.Sf

## I. INTRODUCTION

In recent years, there has been increasing interest in studying solutions of fields with different spins coupled to gravity. The most extensively studied cases involve fields with spin 0 [1, 2] and 1 [3–5], and more recently spinor fields with spin 1/2. In particular, self-gravitating configurations arising from the Einstein–Dirac system in a spherically symmetric spacetime, known in the literature as "Dirac stars", were first investigated in [6]. Due to the intrinsic angular momentum of spinors fields, strictly spherically symmetric configurations cannot be obtained from a single spinor field. Instead, such solutions are constructed from pairs of spinors with opposite spin orientations. Building on this setup, numerous studies have investigated different Dirac star configurations. These include, for instance, $\kappa$-pairs of spinors with different azimuthal indices [7], as well as multi-state configurations involving pairs of Dirac fields in the ground and excited states [8]. Extensions including self-interacting terms have been studied in [9], while nonlinear self-interactions of the spinor field have also been considered in [10].

However, most of these studies neglect electromagnetic interactions. Charged Dirac configurations arise naturally within the Einstein-Dirac-Maxwell framework, where particle-like solutions were first obtained in a spherically symmetric spacetime in [11]. Nevertheless, to the best of our knowledge only very few works have considered a charged Dirac field coupled to gravity [12, 13]. While these approaches provide valuable insight into the existence and properties of such objects, they are limited when addressing dynamical scenarios.

In this context, the 3+1 formulation of general relativity offers a natural framework to study the time evolution of self-gravitating spinor fields. This formalism allows one to cast the Einstein–Dirac–Maxwell system as an initial value problem, making it particularly suitable for numerical implementations, and for exploring dynamical regimes beyond equilibrium configurations. In this work, we develop the $3 + 1$ decomposition of the Einstein–Dirac–Maxwell system in a spherically symmetric spacetime, and construct stationary solutions as a first step toward a fully dynamical treatment to be presented elsewhere.

This paper is organized as follows. In section II we obtain the field equations for the Einstein-Dirac-Maxwell system and rewrite the system using the $3 + 1$ formulation. In section III we consider the particular case of a spherically-symmetric spacetime. Further, in section IV we consider the case of stationary static solutions, and derive the system of equations that corresponds to charged Dirac stars. In section V we present the boundary conditions required to solve the system numerically. In Section VI we describe how to calculate the total mass, charge and binding energy of our solutions. In Section VII we present our numerical results for the different families of charged configurations, and in section VIII we show some particular solutions of the system. Finally, we conclude in section IX.

* maribel.hernandez@nucleares.unam.mx
† malcubi@nucleares.unam.mx

Throughout this paper we use Planck units such that $G = c = \hbar = 1$, and the signature of the spacetime metric is taken to be $(-,+,+,+)$.

## II. THE EINSTEIN–DIRAC–MAXWELL SYSTEM

### A. Action and field equations

We consider a charged Dirac field coupled to a real Maxwell field, both minimally coupled to gravity. The action that describes our system is given by:

$$S = \int \left[ \frac{R}{16\pi} + \mathcal{L}_M + \mathcal{L}_D \right] \sqrt{-g}\, d^4x \; , \tag{1}$$

where $R$ is the Ricci scalar, and where $\mathcal{L}_M$ and $\mathcal{L}_D$ are the Lagrangians of the Maxwell and Dirac fields, respectively. For the Maxwell field we have the standard lagrangian:

$$\mathcal{L}_M = -\frac{1}{16\pi} F^{\mu\nu} F_{\mu\nu} \; , \tag{2}$$

with the Maxwell (Faraday) field tensor $F_{\mu\nu}$ is defined in terms of the potential 1-form $A_\mu$ in the usual way, $F_{\mu\nu} := \nabla_\mu A_\nu - \nabla_\nu A_\mu$. On the other hand, the Lagrangian for the Dirac field coupled to the Maxwell field is given by:

$$\mathcal{L}_D = \frac{i}{2} \left[ \bar{\psi}\gamma^\mu \left( \mathcal{D}_\mu \psi - iqA_\mu \psi \right) - \left( \mathcal{D}_\mu \bar{\psi} + iqA_\mu \bar{\psi} \right) \gamma^\mu \psi \right] - m\bar{\psi}\psi \; , \tag{3}$$

with $\psi$ the Dirac spinor, $\bar{\psi}$ its adjunct, $m$ the mass of the associated fermion, $q$ a charge parameter. In the previous expression $\mathcal{D}_\mu \psi := \partial_\mu \psi + \Gamma_\mu \psi$ and $\mathcal{D}_\mu \bar{\psi} := \partial_\mu \bar{\psi} - \bar{\psi}\Gamma_\mu$ denote the spinor covariant derivatives, with $\Gamma_\mu$ the Fock–Ivanenko coefficients (spinor affine connection) and $\gamma^\mu := \gamma^A e^\mu_A$ the covariant Dirac matrices, with $\gamma^A$ the usual flat space Dirac matrices and $e^\mu_A$ the coordinate basis components of the tetrad $\{\vec{e}_A\}$. In these expressions both the Greek and uppercase Latin indices take values from 0 to 4, with Greek indices representing spacetime components and uppercase Latin indices tetrad components. Furthermore, the tetrad satisfies the following relations:

$$\vec{e}_A \cdot \vec{e}_B = g_{\mu\nu} e^\mu_A e^\nu_B = \eta_{AB} \; . \tag{4}$$

On the other hand, the Fock–Ivanenko coefficients are given in terms of the Christoffel symbols and tetrad components as:

$$\Gamma_\mu = -\frac{1}{4} \omega_{AB\mu} \gamma^A \gamma^B \; , \tag{5}$$

with $\omega_{AB\mu}$ the Ricci rotation coefficients given by:

$$\omega_{AB\mu} = e^\nu_B \left( e_{\lambda A} \Gamma^\lambda_{\nu\mu} - \partial_\mu e_{\nu A} \right) \; , \tag{6}$$

where now $\Gamma^\lambda_{\nu\mu}$ are the usual Christoffel symbols.

Notice that the action above is invariant under the following gauge transformation:

$$\psi \to e^{iq\Lambda}\psi \; , \qquad \bar{\psi} \to e^{-iq\Lambda}\bar{\psi} \; , \qquad A_\mu \to A_\mu + \partial_\mu \Lambda \; , \tag{7}$$

with $\Lambda(x)$ an arbitrary scalar function. This implies the existence of a conserved Noether current given by:

$$j^\mu = q\bar{\psi}\gamma^\mu\psi \; . \tag{8}$$

If we now vary the action with respect to the spacetime metric $g_{\mu\nu}$, the vector potential $A_\mu$, the Dirac field $\psi$, and the tetrad $\{\vec{e}_A\}$, we obtain the Einstein, Dirac and Maxwell field equations:

$$R_{\mu\nu} - \frac{1}{2} g_{\mu\nu} R = 8\pi T_{\mu\nu} \; , \tag{9}$$

$$i(\gamma^\mu \mathcal{D}_\mu \psi - iq\gamma^\mu A_\mu \psi) - m\psi = 0 \; , \tag{10}$$

$$\nabla_\nu F^{\nu\mu} = -4\pi j^\mu \; , \tag{11}$$

with $j^\mu$ the conserved current defined above, and where the total stress-energy tensor includes contributions from both the Dirac and Maxwell fields, $T_{\mu\nu} = T^{(D)}_{\mu\nu} + T^{(M)}_{\mu\nu}$, with:

$$\begin{aligned} T^{(D)}_{\mu\nu} &= \frac{i}{2}\left[\mathcal{D}_{(\mu}\bar{\psi}\gamma_{\nu)}\psi + iq\bar{\psi}\gamma_\nu A_\mu\psi - \left(\bar{\psi}\gamma_{(\mu}\mathcal{D}_{\nu)}\psi - iq\bar{\psi}\gamma_\mu A_\nu\psi\right)\right] \\ &= \frac{i}{2}\left[\mathcal{D}_{(\mu}\bar{\psi}\gamma_{\nu)}\psi - \bar{\psi}\gamma_{(\mu}\mathcal{D}_{\nu)}\psi + iq\bar{\psi}\left(\gamma_\nu A_\mu + \gamma_\mu A_\nu\right)\psi\right] , \end{aligned} \tag{12}$$

$$T^{(M)}_{\mu\nu} = \frac{1}{4\pi}\left(F_{\mu\alpha}F_\nu{}^\alpha - \frac{g_{\mu\nu}}{4}\,F_{\alpha\beta}F^{\alpha\beta}\right) . \tag{13}$$

### B. The Maxwell equations in formalism 3+1

In this section we rewrite the Einstein–Dirac–Maxwell system of equations using the 3+1 formalism of general relativity, where the spacetime $(\mathcal{M}, g_{\mu\nu})$ is foliated by Cauchy hypersurfaces $\Sigma_t$ parametrized by a global time function $t$, resulting in the standard Arnowitt–Deser–Misner (ADM) equations [14]. The 3+1 split of the metric takes the following form:

$$ds^2 = (-\alpha^2 dt^2 + \beta_i\beta^i)dt^2 + 2\beta_i dt dx^i + \gamma_{ij}dx^i dx^j , \tag{14}$$

where $\{x^i\}$ are spatial coordinates, $\gamma_{ij}$ is the induced three-dimensional metric defined within the spatial hypersurfaces, $\alpha$ is the lapse function, and $\beta^i$ is the shift vector (with $\beta_i = \gamma_{ij}\beta^j$) [14]. The unit normal vector $n^\mu$ to the spatial hypersurfaces has the following components:

$$n^\mu = (1/\alpha, -\beta^i/\alpha) , \qquad n_\mu = (-\alpha, 0, 0, 0) . \tag{15}$$

We can decompose any tensor field into its components normal and parallel to the spatial hypersurfaces by projecting it with the normal unit vector $n^\mu$ and the projector operator $P^\mu_\nu := \delta^\nu_\mu + n^\nu n_\mu \equiv \gamma^\mu_\nu$, respectively. From the Maxwell potential 1-form $A_\mu$ we define the scalar and 3-vector potentials as:

$$\phi := -n^\mu A_\mu , \qquad a_\mu := P^\nu_\mu A_\nu . \tag{16}$$

In particular, these definitions imply that $n^\mu a_\mu = 0$, that is $a_\mu$ is a purely spatial vector, so from now on we will only consider its spatial components. Similarly, from the Faraday field tensor $F_{\mu\nu}$ we can define its "electric" and "magnetic" parts as:

$$E_\mu := n^\nu F_{\mu\nu} , \qquad B_\mu := n^\nu F^*_{\mu\nu} , \tag{17}$$

where $F^*_{\mu\nu}$ is the dual Faraday tensor given by $F^*_{\alpha\beta} := -\frac{1}{2}\,\tilde{\varepsilon}_{\alpha\beta\mu\nu}F^{\mu\nu}$, with $\tilde{\varepsilon}_{\alpha\beta\mu\nu}$ the Levi–Civita tensor which is in turn defined in terms of the Levi–Civita symbol $\varepsilon_{\alpha\beta\mu\nu}$ as $\tilde{\varepsilon}_{\alpha\beta\mu\nu} := \sqrt{-g}\,\varepsilon_{\alpha\beta\mu\nu}$ (we use a convention such that $\epsilon_{0123} = 1 = -\epsilon^{0123}$). Again, one can easily see that $n^\mu E_\mu = n^\mu B_\mu = 0$, so we only need to consider their spatial components.

The energy density, momentum density, and spatial stress tensor for the Maxwell field as measured by the Eulerian observers (those moving along the normal direction to the spatial hypersurfaces) are defined in the usual way as:

$$\rho_M := n^\mu n^\nu T^{(M)}_{\mu\nu} = \frac{a^2}{8\pi}\left(B^2 + E^2\right) , \tag{18}$$

$$J^i := -n^\mu P i^\nu T^{(M)}_{\mu\nu} = \frac{1}{4\pi}\left(E \times B\right)^i , \tag{19}$$

$$S_{ij} := P^\mu_i P^\nu_j T^{(M)}_{\mu\nu} = \frac{1}{8\pi}\left[\gamma_{ij}\left(E^2 + B^2\right) - 2\left(B_iB_j + E_iE_j\right)\right] , \tag{20}$$

where $E^2 = E_iE^i$, $B^2 = B_iB^i$, and $(E \times B)^i = \tilde{\varepsilon}^{ijk}E_jB_k$, with $\tilde{\varepsilon}^{ijk} := n_\mu\tilde{\varepsilon}^{\mu ijk}$ the three-dimensional Levi–Civita tensor for the spatial hypersurfaces.

If we now impose the Lorenz gauge condition $\nabla_\nu A^\nu = 0$, we can obtain an evolution equation for the scalar potential $\phi$, given by:

$$\partial_t\phi - \mathcal{L}_{\vec{\beta}}\phi = -D_i(\alpha a^i) + \alpha\phi K , \tag{21}$$

where here $D_i$ is the standard three-dimensional covariant derivative associated with the spatial metric $\gamma_{ij}$. From the definition of the electric field $E_i = n^\nu F_{i\nu}$ we can also find and evolution equation for the vector potential $a_i$:

$$\partial_t a_i - \mathcal{L}_{\vec{\beta}} a_i = -\alpha E_i - \partial_i(\alpha\phi) \; . \tag{22}$$

On the other hand, if we project Maxwell's equations (11) onto the spatial hypersurfaces, we find the following evolution equation for the electric field:

$$\partial_t E^i - \mathcal{L}_{\vec{\beta}} E^i = \alpha K E^i + \left[D \times (\alpha B)^i\right] - 4\pi\alpha q \bar{\psi}\lambda^i\psi \; , \tag{23}$$

with the curl operator for an arbitrary 3-vector $v^i$ is now defined as $(D \times v)^i := \tilde{\varepsilon}^{ijk} D_j v_k$.

Finally, if we project the homogeneous Maxwell equations $\nabla_\mu F^{\mu\nu *} = 0$ onto the spatial hypersurfaces we find an evolution equation for the magnetic field:

$$\partial_t B^i - \mathcal{L}_{\vec{\beta}} B^i = \alpha K B^i - \left[D \times (\alpha E)\right]^i \; . \tag{24}$$

Additionally, if we contract Maxwell's equations with the normal vector $n^\mu$ we find the Gauss and magnetic constraints given by:

$$D_i E^i \;=\; 4\pi q \psi^\dagger \psi = 4\pi q \sum_i^4 |\psi_i|^2 \; , \tag{25}$$

$$D_i B^i \;=\; 0 \; . \tag{26}$$

### C. The Dirac equation in 3+1 form

In order to rewrite the Dirac equation (10) in 3+1 form we will follow closely the discussion found in reference [15]. We first have to choose a tetrad adapted to our spacetime foliation, we will take:

$$e^\mu_T = n^\mu = (1/\alpha, -\beta^i/\alpha) \; , \qquad e^\mu_I = E^\mu_I \; , \tag{27}$$

where $E^\mu_I$ ($I \in \{1,2,3\}$) are three purely spatial unitary vectors orthogonal to each other, which we will from now on call the "spatial triad". Here it is important to notice that we have four different type of indices: 1) spacetime coordinate indices that take values from 0 to 3 for which we use the Greek letters $\{\alpha, \beta ...\}$; 2) tetrad indices that take values from 0 to 3, for which we use upper case Latin letters starting from $\{A, B, ,..\}$; 3) purely spatial coordinates that take values from 1 to 3, for which we use lower case Latin letters starting from $\{i, j, ..\}$; and 4) purely spatial triad indices that take values from 1 to 3, and for which we use upper case Latin indices starting from $\{I, J, ..\}$.

The Dirac equation (10) can now be written as:

$$i[\gamma^\mu(\partial_\mu\psi + \Gamma_\mu\psi) - iq\gamma^\mu A_\mu \psi] - m\psi = 0 \; , \tag{28}$$

or equivalently:

$$i[\gamma^t\partial_t\psi + \gamma^m\partial_m\psi + \gamma^\mu\Gamma_\mu\psi - iq\gamma^\mu A_\mu\psi] - m\psi = 0 \; . \tag{29}$$

The $\gamma^\mu$ matrices can be written in terms of the tetrad as:

$$\gamma^t \;=\; e^t_A \gamma^A = e^t_T \gamma^T + e^t_I \gamma^I = \left(\frac{1}{\alpha}\right)\gamma^T \; , \tag{30}$$

$$\gamma^m \;=\; e^m_A \gamma^A = e^m_T \gamma^T + e^m_I \gamma^I = -\left(\frac{\beta^m}{\alpha}\right)\gamma^T + E^m_I \gamma^I = -\left(\frac{\beta^m}{\alpha}\right)\gamma^T + \lambda^m \; , \tag{31}$$

where we have defined the purely spatial Dirac matrices as $\lambda^m := E^m_I \gamma^I$. Replacing now (30) and (31) into equation (29), and multiplying by $i$, we find:

$$\frac{\gamma^T}{\alpha}\left(\partial_t\psi - \beta^m\partial_m\psi\right) + \lambda^m\partial_m\psi + \gamma^\mu\Gamma_\mu\psi - iq\gamma^\mu A_\mu\psi + im\psi = 0 \; . \tag{32}$$

Multiplying this last equation with $\alpha\gamma^T$ from the left we obtain:

$$\left(\partial_t - \beta^m\partial_m\right)\psi + \alpha\gamma^T\left[\lambda^m\partial_m + \gamma^\mu\Gamma_\mu - iq\gamma^\mu A_\mu + im\right]\psi = 0 \; . \tag{33}$$

Also, according to equation (16) we have:

$$\phi = -n^\nu A_\nu = -\frac{1}{\alpha}\left(A_t - \beta^m A_m\right) \; , \tag{34}$$

and since $A_\mu = a_\mu + n_\mu \phi$ we find $A_m = a_m$, so that:

$$\begin{aligned} \gamma^T \left( i q \gamma^\mu A_\mu \right) &= i q \gamma^T \left( \gamma^t A_t + \gamma^m A_m \right) \\ &= i q \gamma^T \left[ \frac{\gamma^T}{\alpha} \left( A_t - \beta^m A_m \right) + \lambda^m A_m \right] \\ &= i q \left( -\phi + \gamma^T \lambda^m a_m \right) \; . \end{aligned} \tag{35}$$

Furthermore, it can be shown that:

$$\gamma^\mu \Gamma_\mu = \left( \frac{\partial_I \alpha}{2\alpha} \right) \gamma^I - \frac{1}{4} Q_{IJ} \gamma^T \gamma^I \gamma^J - \frac{K}{2} \, \gamma^T + \gamma^I \Gamma^{(3)}_I \; , \tag{36}$$

where $Q_{IJ} := \omega_{IJT}$, and with $K$ the trace of the extrinsic curvature $K_{mn}$ of the spatial hypersurfaces of constant time, $K = K^m{}_m$. As discussed in detail in reference [15], the quantities $Q_{IJ}$ are antisymmetric and represent our freedom in choosing how the spatial triad propagates through time. For a triad that is propagated by Fermi–Walker transport (i.e. following the local inertial frame of the Eulerian observers without rotating) we will always have $Q_{IJ} = 0$.

If we now multiply equation (36) with $\gamma^T$ from the left we find:

$$\begin{aligned} \gamma^T \left( \gamma^\mu \Gamma_\mu \right) &= \gamma^T \left( \frac{\partial_I \alpha}{2\alpha} \right) \gamma^I - \frac{1}{4} \, Q_{IJ} \gamma^I \gamma^J - \frac{K}{2} + \gamma^T \gamma^I \Gamma^{(3)}_I \\ &= \gamma^T \lambda^m \left( \frac{\partial_m \alpha}{2\alpha} + \Gamma^{(3)}_m \right) - \frac{1}{4} \, Q_{IJ} \gamma^I \gamma^J - \frac{K}{2} \; , \end{aligned} \tag{37}$$

where in the last equality we used the fact that $\gamma^I \partial_I = \lambda^m \partial_m$ and $\gamma^I \Gamma^{(3)}_I = \lambda^m \Gamma^{(3)}_m$. Substituting now equations (35) and (37) into (33), we obtain:

$$\left( \partial_t - \beta^m \partial_m \right) \psi = \alpha \left[ -\gamma^T \lambda^m \left( \partial_m + {}^{(3)}\Gamma_m + \frac{\partial_m \alpha}{2\alpha} - i q a_m \right) + \left( \frac{K}{2} - i q \phi - i m \gamma^T \right) + \frac{1}{4} \, Q_{IJ} \lambda^I \lambda^J \right] \psi \; , \tag{38}$$

and defining the three-dimensional spinor covariant derivative as $D_m \psi := \partial_m \psi + \Gamma^{(3)}_m \psi$, we finally find:

$$\left( \partial_t - \beta^m \partial_m \right) \psi = -\alpha \gamma^T \left[ \lambda^m \left( D_m + \frac{\partial_m \alpha}{2\alpha} - i q a_m \right) + i m \right] \psi + \alpha \left( \frac{K}{2} - i q \phi + \frac{1}{4} \, Q_{mn} \lambda^m \lambda^n \right) \psi \; . \tag{39}$$

On the other hand, from equation (5) we have:

$$\Gamma_T := -\frac{1}{4} \, \omega_{ABT} \gamma^A \gamma^B \; , \tag{40}$$

which can be shown to imply:

$$\Gamma_T = \gamma^T \lambda_m \left( \frac{\partial_m \alpha}{2\alpha} \right) - \frac{1}{4} Q_{mn} \lambda^m \lambda^n \; . \tag{41}$$

Equation (39) can then be writte in more compact form as:

$$\left( \partial_t - \beta^m \partial_m \right) \psi = -\alpha \gamma^T \left[ \lambda^m \left( D_m - i q a_m \right) + i m \right] \psi + \alpha \left( \frac{K}{2} - i q \phi - \Gamma_T \right) \psi \; . \tag{42}$$

This is the final form of the charged Dirac equation in the 3+1 formalism.

### D. Conserved current and stress-energy tensor for the Dirac field in the 3+1 form

The conserved Noether current for the Dirac field is be given by equation (8), $j^\mu = q\bar{\psi}\gamma^\mu\psi$. As usual, we can define the charge density $\rho_q$ measured by the Eulerian observers as $\rho_q := -n_\mu j^\mu$, we then find:

$$\rho_q = q\ \left(|\psi_1|^2 + |\psi_2|^2 + |\psi_3|^2 + |\psi_4|^2\right)\ , \tag{43}$$

where $\psi_i$, with $i = 1,2,3,4$, are the four components of the Dirac spinor.

In a similar way, the electric current measured by the Eulerian observers is defined as $f^i := P^i_\nu j^\nu$, with $P^\mu_\nu$ the projection operator onto the spatial hypersurfaces defined before. We then have:

$$f^i = q\,\bar{\psi}\lambda^i\psi\ . \tag{44}$$

On the other hand, the energy density measured by Eulerian observers is now defined as:

$$\begin{aligned}
\rho_D &:= n^\mu n^\nu T^{(D)}_{\mu\nu} \\
&= \frac{i}{2}\,n^\mu n^\nu\left[\mathcal{D}_{(\mu}\bar{\psi}\gamma_{\nu)}\psi - \bar{\psi}\gamma_{(\mu}\mathcal{D}_{\nu)}\psi\right] + \frac{i}{2}n^\mu n^\nu\left[iq\bar{\psi}\gamma_\nu A_\mu\psi + iq\bar{\psi}\gamma_\mu A_\nu\psi\right] \\
&= \frac{i}{2}\left[\psi^\dagger\Pi - \Pi^\dagger\psi\right] - q\phi\sum_{i=1}^{4}|\psi_i|^2\ ,
\end{aligned} \tag{45}$$

where we have used the definitions $\Pi := n^\mu\mathcal{D}_\mu\psi$ and $\bar{\Pi} := n^\mu\mathcal{D}_\mu\bar{\psi}$, which can be rewritten as:

$$\Pi = \frac{1}{\alpha}\left(\partial_t\psi - \beta^i\partial_i\psi\right) + \Gamma_T\psi\ , \qquad \bar{\Pi} = \frac{1}{\alpha}\left(\partial_t\bar{\psi} - \beta^i\partial_i\bar{\psi}\right) - \bar{\psi}\Gamma_T\ . \tag{46}$$

Alternatively, we can also define $\tilde{\Pi} := n^\mu\partial_\mu\psi = \frac{1}{\alpha}(\partial_t\psi - \beta^i\partial_i\psi)$ and $\tilde{\Pi}^\dagger := n^\mu\partial_\mu\psi^\dagger = \frac{1}{\alpha}(\partial_t\psi^\dagger - \beta^i\partial_i\psi^\dagger)$, so that:

$$\Pi = \tilde{\Pi} + \Gamma_T\psi\ , \qquad \Pi^\dagger = \tilde{\Pi}^\dagger + \psi^\dagger\Gamma_T^\dagger\ . \tag{47}$$

In that case the expression for the energy density $\rho_D$ (45) becomes:

$$\begin{aligned}
\rho_D &= \frac{i}{2}\left[\psi^\dagger\Pi - \Pi^\dagger\psi\right] - q\phi\sum_{i=1}^{4}|\psi_i|^2 \\
&= \frac{i}{2}\left[\psi^\dagger\tilde{\Pi} - \tilde{\Pi}^\dagger\psi + \psi^\dagger\left(\Gamma_T - \Gamma_T^\dagger\right)\psi\right] - q\phi\sum_{i}^{4}|\psi_i|^2 \\
&= \frac{i}{2}\left[\psi^\dagger\tilde{\Pi} - \tilde{\Pi}^\dagger\psi - \frac{1}{2}\psi^\dagger\left(Q_{mn}\lambda^m\lambda^n\right)\psi\right] - q\phi\sum_{i=1}^{4}|\psi_i|^2,
\end{aligned} \tag{48}$$

where we have used equation (41).

We can also compute the momentum density measured by the Eulerian observers defined as $\mathcal{J}_i := -n^\mu P^\nu_i T^{(D)}_{\mu\nu}$. We find:

$$\begin{aligned}
\mathcal{J}_i &= -\frac{i}{2}\,n^\mu P^\nu_i\left[\mathcal{D}_{(\mu}\bar{\psi}\gamma_{\nu)}\psi - \bar{\psi}\gamma_{(\mu}\mathcal{D}_{\nu)}\psi\right] + \frac{q}{2}\,n^\mu P^\nu_i\left[\bar{\psi}\gamma_\nu A_\mu\psi + \bar{\psi}\gamma_\mu A_\nu\psi\right] \\
&= -\frac{i}{4}\left[\bar{\Pi}\lambda_i\psi - \bar{\psi}\lambda_i\Pi + \psi^\dagger(D_i\psi) - (D_i\psi^\dagger)\psi\right] - \frac{q}{2}\,\bar{\psi}\left[\gamma^T a_i + \lambda_i\phi\right]\psi\ ,
\end{aligned} \tag{49}$$

where we have used the fact that the first term in $\mathcal{J}_i$ is the same that is found when there is no charge (see reference [15]), while for the second term we have $n^\mu A_\mu = -\phi$, $P^\nu_i\gamma_\nu = \lambda_i$, $\gamma^T = -n^\mu\gamma_\mu$, and $P^\nu_i A_\nu = a_i$. In the above expression one should note that we have two different types of spinor derivatives (see reference[15]): the usual 4-dimensional spinor derivative $\mathcal{D}_\mu\psi = \partial_\mu\psi + \Gamma_\mu\psi$, and the 3-dimensional spinor derivative $D_i\psi = \partial_i\psi + \Gamma^{(3)}_i\psi$, where the $\Gamma^{(3)}_i$ are the 3-dimensional Fock–Ivanenko coefficients which are given by:

$$\Gamma^{(3)}_i = -\frac{1}{4}\,\omega^{(3)}_{imn}\gamma^m\gamma^n = \Gamma_i + \frac{1}{2}\,K_{im}\gamma^T\gamma^m\ , \tag{50}$$

so that:

$$P_i^\nu \mathcal{D}_\nu \psi = \mathcal{D}_i \psi = D_i \psi + \frac{1}{2}\, K_{im} \gamma^T \gamma^m \psi \; . \tag{51}$$

Finally, we can compute the components of the spatial stress tensor $\mathcal{S}_{ij}$:

$$\begin{aligned} \mathcal{S}_{ij} &:= P_i^\mu P_j^\nu T^{(D)}_{\mu\nu} \\ &= \frac{i}{2}\, P_i^\mu P_i^\nu \left[ \mathcal{D}_{(\mu} \bar{\psi} \gamma_{\nu)} \psi - \bar{\psi} \gamma_{(\mu} \mathcal{D}_{\nu)} \psi \right] + \frac{q}{2}\, P_i^\mu P_i^\nu \bar{\psi} \left[ \gamma_\nu A_\mu + \gamma_\mu A_\nu \right] \psi \; . \end{aligned} \tag{52}$$

For the first term above we can use the result previously found in reference [15], while for the second term we use equation (16). We finally find:

$$\mathcal{S}_{ij} = \frac{i}{2} \left[ \left( D_{(i} \bar{\psi} \right) \lambda_{j)} \psi - \bar{\psi} \lambda_{(i} \left( D_{j)} \psi \right) - \psi^\dagger \left( K_{ij} + K_{m(i} \lambda_{j)} \lambda^m \right) \psi \right] - \frac{q}{2}\, \bar{\psi} \left[ \lambda_i a_j + \lambda_j a_i \right] \psi \; . \tag{53}$$

In particular, for the trace of $S_{ij}$ we find:

$$S \equiv S^m{}_m \;=\; \frac{i}{2} \left[ \left( D_m \bar{\psi} \right) \lambda^m \psi - \bar{\psi} \lambda^m \left( D_m \psi \right) \right] - q \bar{\psi} \left( \lambda^m a_m \right) \psi \; , \tag{54}$$

where we used the fact that $K_{mn} \lambda^m \lambda^n = -K$.

## III. SPHERICALLY SYMMETRIC SPACETIME

We will now consider the particular case of a spherically-symmetric spacetime in spherical coordinates $(t, r, \theta, \varphi)$. The metric can be written in general as:

$$ds^2 = (-\alpha^2 + \beta_r \beta^r) dt^2 + 2\beta_r dr dt + \mathrm{a}^2 dr^2 + b^2 r^2 d\Omega^2, \tag{55}$$

where $d\Omega^2 = d\theta^2 + \sin^2\theta d\varphi^2$ is the standard solid angle element, $\alpha = \alpha(r, t)$ is the lapse function, $\beta^r = \beta^r(r, t)$ is the radial component of the shift vector, and $\mathrm{a} = \mathrm{a}(r, t)$ and $b = b(r, t)$ are the spatial metric components.

In order to write the Dirac equation we choose the following tetrad:

$$e^\mu_T = n^\mu = \left( \frac{1}{\alpha}, -\frac{\beta^r}{\alpha}, 0, 0 \right) \;, \quad e^\mu_R = \left( 0, \frac{1}{\mathrm{a}}, 0, 0 \right) \;, \quad e^\mu_\Theta = \left( 0, 0, \frac{1}{rb}, 0 \right) \;, \quad e^\mu_\Phi = \left( 0, 0, 0, \frac{1}{rb \sin\theta} \right) \; . \tag{56}$$

In order to avoid confusions from here on we will denote the spacetime indices by $(t, r, \theta, \varphi)$ and their associated tetrad indices by $(T, R, \Theta, \Phi)$. The spatial triad then takes the form:

$$E^i_R = \left( \frac{1}{\mathrm{a}}, 0, 0 \right) \;, \qquad E^i_\Theta = \left( 0, \frac{1}{rb}, 0 \right) \;, \qquad E^i_\Phi = \left( 0, 0, \frac{1}{rb \sin\theta} \right) \; . \tag{57}$$

Following [15], we will take for the $\gamma^A$ matrices the following convention:

$$\gamma^T = \gamma^0 \;, \qquad \gamma^R = \gamma^3 \;, \qquad \gamma^\Theta = \gamma^2 \;, \qquad \gamma^\Phi = \gamma^1 \; . \tag{58}$$

For their spacetime components $\gamma^\mu = e^\mu_A \gamma^A$ we then have:

$$\gamma^t = \frac{\gamma^T}{\alpha} \;, \qquad \gamma^r = \frac{\gamma^R}{\mathrm{a}} - \frac{\beta^r \gamma^T}{\alpha} \;, \qquad \gamma^\theta = \frac{\gamma^\Theta}{rb} \;, \qquad \gamma^\varphi = \frac{\gamma^\Phi}{rb \sin\theta} \; . \tag{59}$$

On the other hand, for the purely spatial Dirac matrices $\lambda$ we will have:

$$\lambda^R = \gamma^3 \;, \qquad \lambda^\Theta = \gamma^2 \;, \qquad \lambda^\Phi = \gamma^1 \;, \tag{60}$$

and:

$$\lambda^r = \frac{\gamma^R}{\mathrm{a}} \;, \qquad \lambda^\theta = \frac{\gamma^\Theta}{rb} \;, \qquad \lambda^\varphi = \frac{\gamma^\Phi}{rb \sin\theta} \; . \tag{61}$$

Since we are considering spherical symmetry our spatial triad does not rotate, which implies $Q_{mn} = 0$. The Dirac equation then reduces to:

$$(\partial_t - \beta^r \partial_r)\,\psi = -\alpha\gamma^T \left[\lambda^r \left(\partial_r + \frac{\partial_r \alpha}{2\alpha}\right) + \lambda^m \Gamma^{(3)}_m + im\right]\psi + \left(\frac{\alpha K}{2} - iq\phi\right)\psi \; . \tag{62}$$

Using the above expressions for the $\lambda^m$, and the expression for $\lambda^I \Gamma^{(3)}_I = \lambda^m \Gamma^{(3)}_m$ found in [15], we find for each of the spinor components:

$$\begin{aligned}
(\partial_t - \beta^r \partial_r)\psi_1 \;&=\; \alpha \left[\frac{-1}{a}\left(\partial_r + \frac{\partial_r \alpha}{2\alpha} + \frac{\partial_r b}{b} + \frac{1}{r} - iqa_r\right)\psi_3 + \frac{i}{rb}\left(\partial_\theta + \frac{i\partial_\varphi}{\sin\theta} + \frac{\cot\theta}{2} + \frac{qa_\varphi}{\sin\theta} - iqa_\theta\right)\psi_4\right] \\
&+\; \alpha\left(\frac{K}{2} - im - iq\phi\right)\psi_1 \; , \qquad (63) \\
(\partial_t - \beta^r \partial_r)\psi_2 \;&=\; \alpha \left[\frac{1}{a}\left(\partial_r + \frac{\partial_r \alpha}{2\alpha} + \frac{\partial_r b}{b} + \frac{1}{r} - iqa_r\right)\psi_4 - \frac{i}{rb}\left(\partial_\theta - \frac{i\partial_\varphi}{\sin\theta} + \frac{\cot\theta}{2} - \frac{qa_\varphi}{\sin\theta} - iqa_\theta\right)\psi_3\right] \\
&+\; \alpha\left(\frac{K}{2} - im - iq\phi\right)\psi_2 \; , \qquad (64) \\
(\partial_t - \beta^r \partial_r)\psi_3 \;&=\; \alpha \left[-\frac{1}{a}\left(\partial_r + \frac{\partial_r \alpha}{2\alpha} + \frac{\partial_r b}{b} + \frac{1}{r} - iqa_r\right)\psi_1 + \frac{i}{rb}\left(\partial_\theta + \frac{i\partial_\varphi}{\sin\theta} + \frac{\cot\theta}{2} + \frac{qa_\varphi}{\sin\theta} - iqa_\theta\right)\psi_2\right] \\
&+\; \alpha\left(\frac{K}{2} + im - iq\phi\right)\psi_3 \; , \qquad (65) \\
(\partial_t - \beta^r \partial_r)\psi_4 \;&=\; \alpha \left[\frac{1}{a}\left(\partial_r + \frac{\partial_r \alpha}{2\alpha} + \frac{\partial_r b}{b} + \frac{1}{r} - iqa_r\right)\psi_2 - \frac{i}{rb}\left(\partial_\theta - \frac{i\partial_\varphi}{\sin\theta} + \frac{\cot\theta}{2} - \frac{qa_\varphi}{\sin\theta} - iqa_\theta\right)\psi_1\right] \\
&+\; \alpha\left(\frac{K}{2} + im - iq\phi\right)\psi_4 \; . \qquad (66)
\end{aligned}$$

In order to solve the above system we propose the ansatz $\psi_i = R_i(t,r)T_i(\theta,\varphi)$, where $i = 1,2,3,4$ denotes spinor components, and $(R_i, T_i)$ are complex functions to be determined. From the above equations one can show that this system is separable if we take $a_\theta = a_\varphi = 0$, with $\phi = \phi(r,t)$, $a_r = a_r(r,t)$ functions of $r$ and $t$ only. One also finds that we need to take $R_2 = iR_1$, $R_4 = iR_3$, $T_3 = T_1$, $T_4 = -T_2$, so that we are only left with two functions of $(r,t)$ and two functions of $(\theta,\varphi)$ to be determined. Renaming the radial functions as $R_1(t,r) \equiv F(t,r)$ and $R_3(t,r) \equiv G(t,r)$, our system of radial equations reduces to:

$$\frac{1}{\alpha}\left(\partial_t F - \beta^r \partial_r F\right) - \left(\frac{K}{2} - iq\phi - im\right)F + \frac{1}{a}\left(\partial_r + \frac{\partial_r \alpha}{2\alpha} + \frac{\partial_r b}{b} + \frac{1}{r} - iqa_r\right)G = k\,\frac{G}{rb} \; , \tag{67}$$

$$\frac{1}{\alpha}\left(\partial_t G - \beta^r \partial_r G\right) - \left(\frac{K}{2} - iq\phi + im\right)G + \frac{1}{a}\left(\partial_r + \frac{\partial_r \alpha}{2\alpha} + \frac{\partial_r b}{b} + \frac{1}{r} - iqa_r\right)F = -k\,\frac{F}{rb} \; , \tag{68}$$

while the equations for the angular part are:

$$\left(\partial_\theta - \frac{i}{\sin\theta} + \frac{\cot\theta}{2}\right)T_1 \;=\; -kT_2 \; , \tag{69}$$

$$\left(\partial_\theta + i\frac{\partial_\varphi}{\sin\theta} + \frac{\cot\theta}{2}\right)T_2 \;=\; kT_1 \; , \tag{70}$$

with $k$ a separation constant. Taking $k = -1$ the solutions for the angular functions are $T_1 \equiv T_+$ and $T_2 \equiv T_-$, where:

$$T_\pm = \pm\left(\frac{1}{\sqrt{4\pi}}\right)e^{\pm i\varphi/2}y_\pm(\theta) \; , \tag{71}$$

with $y_+(\theta) = \sin(\theta/2)$ and $y_- = \cos(\theta/2)$.

The spinors associated with our solutions will then be given by:

$$\psi_\pm \;=\; \frac{e^{\pm i\varphi/2}}{\sqrt{4\pi}}\begin{pmatrix} F(t,r)y_\pm\,(\theta) \\ \pm iF(t,r)\,y_\mp(\theta) \\ G(t,r)\,y_\pm(\theta) \\ \mp iG(t,r)\,y_\mp(\theta) \end{pmatrix} \; . \tag{72}$$

In terms of these functions, and considering contributions from the two spinors, we find for the total charge density and electric current:

$$\rho_q^{(\mathrm{Tot})} = \frac{q}{2\pi}\left(|F|^2+|G|^2\right) , \tag{73}$$

$$f_r^{(\mathrm{Tot})} = \frac{\mathrm{a}}{2\pi}\left(FG^*+F^*G\right) , \tag{74}$$

and for the energy density, momentum density and stress tensor:

$$\begin{aligned}\rho_D^{(\mathrm{Tot})} &= \frac{i}{4\pi}\left[F^*\tilde{\Pi}_F+G^*\tilde{\Pi}_G-c.c\right]-\frac{q\phi}{2\pi}\left(|F|^2+|G|^2\right)\\ &= \frac{1}{2\pi}\left[m\left(|F|^2-|G|^2\right)+\mathrm{Im}\left(\frac{1}{\mathrm{a}}\left(F^*\partial_r G+G^*\partial_r F\right)+\frac{2}{rb}F^*G\right)\right]-\frac{qa_r}{2\pi\mathrm{a}}\left(F^*G+G^*F\right)\end{aligned} \tag{75}$$

$$\begin{aligned}\mathcal{J}_r^{(\mathrm{Tot})} &= \frac{1}{4\pi}\,\mathrm{Im}\left[F^*\partial_r F+G^*\partial_r G-\mathrm{a}\left(F^*\tilde{\Pi}_G+G^*\tilde{\Pi}_F\right)\right]-\frac{q\mathrm{a}}{4\pi}\left[\phi\left(F^*G+G^*F\right)\right]-qa_r\frac{|F|^2+|G|^2}{4\pi}\\ &= \frac{1}{2\pi}\,\mathrm{Im}\left(F^*\partial_r F+G^*\partial_r G\right)-\frac{qa_r}{2\pi}\left(|F|^2+|G|^2\right) ,\end{aligned} \tag{76}$$

$$\mathcal{S}_{rr}^{(\mathrm{Tot})} = \frac{\mathrm{a}}{2\pi}\,\mathrm{Im}\left[F^*\partial_r G+G^*\partial_r F\right]-\frac{q\mathrm{a}a_r}{2\pi}\left(F^*G+G^*F\right) , \tag{77}$$

$$\mathcal{S}_{\theta\theta}^{(\mathrm{Tot})} = \frac{rb}{2\pi}\,\mathrm{Im}\left(F^*G\right) , \tag{78}$$

$$\mathcal{S}_{\varphi\varphi}^{(\mathrm{Tot})} = \frac{rb\sin^2\theta}{2\pi}\,\mathrm{Im}\left(F^*G\right) . \tag{79}$$

## IV. CHARGED DIRAC STARS

Dirac stars correspond to self-gravitating solutions such that the spacetime is static. In that case we can take a vanishing shift vector $\beta^r=0$. Moreover, from now on we will choose the areal radial coordinate, so that the angular metric coefficient is such that $b=1$. The spacetime metric then reduces to:

$$ds^2 = -\alpha(r)^2dt^2+\mathrm{a}^2dr^2+r^2d\Omega^2 , \tag{80}$$

where now $\alpha$ and $a$ are only functions of $r$.

In order to obtain a spherically symmetric energy-momentum tensor we also need to consider equal contributions from the two spinors $\psi_+$ and $\psi_-$ given by in (72). And, since we want a static solution, both $j^\mu$ and $T_{\mu\nu}$ must be time independent and the associated electric current and momentum density have to vanish. In order to achieve this we propose the standard harmonic ansatz:

$$F(r,t)=f(r)e^{-i\omega t} , \qquad G(r,t)=ig(r)e^{-iwt} , \tag{81}$$

with $\omega$ a real frequency to be determined. Also, since we are dealing with an electrostatic problem we will take $a_r=0$.

With these assumptions we now find:

$$\rho_q^{(\mathrm{Tot})} = \frac{q}{2\pi}\left(f^2+g^2\right) , \tag{82}$$

$$f_r^{(\mathrm{Tot})} = 0 , \tag{83}$$

$$\rho_D^{(\mathrm{Tot})} = \frac{1}{2\pi}\left[m\left(f^2-g^2\right)+\frac{1}{\mathrm{a}}\left(f\partial_r g-g\partial_r f\right)+2\,\frac{fg}{rb}\right] , \tag{84}$$

$$\mathcal{J}_r^{(\mathrm{Tot})} = 0 , \tag{85}$$

$$\mathcal{S}_r^{r\,(\mathrm{Tot})} = \frac{1}{2\pi\mathrm{a}}\left(f\partial_r g-g\partial_r f\right) , \tag{86}$$

$$\mathcal{S}_\theta^{\theta\,(\mathrm{Tot})} = \frac{fg}{2\pi r} . \tag{87}$$

Also, from equations (67)-(68) we see that the functions $f,g$ must now satisfy the following ordinary differential

equations:

$$\partial_r f = -f\left(\frac{\partial_r \alpha}{2\alpha} + \frac{1}{r}\left(1 - \mathrm{a}\right)\right) - \mathrm{a} g\left(m + \frac{\omega}{\alpha} - q\phi\right) , \quad (88)$$

$$\partial_r g = -g\left(\frac{\partial_r \alpha}{2\alpha} + \frac{1}{r}\left(1 + \mathrm{a}\right)\right) - \mathrm{a} f\left(m - \frac{\omega}{\alpha} + q\phi\right) , \quad (89)$$

where we have used the fact that the extrinsic curvature vanishes for a static spacetime.

Using the same argument presented in [15], it can be shown that the function $f$ must be an even function of $r$ while the function $g$ must be odd, such that for small $r$ we must have:

$$f \approx f_0 + \mathcal{O}(r^2) , \qquad g \approx g_1 r + \mathcal{O}(r^3) , \quad (90)$$

where $f_0$ and $g_1$ are some constants. Replacing these expansions in (88) one can now show that:

$$\partial_r f|_{r=0} = 0 , \qquad \partial_r g|_{r=0} = g_1 = \left(\frac{\omega}{\alpha} - q\phi - m\right)\frac{f_0}{3} . \quad (91)$$

We still need equations to determine the metric functions $\alpha$ and $\mathrm{a}$. From the Hamiltonian constraint one can immediately find an equation for the radial metric $\mathrm{a}$:

$$\partial_r \mathrm{a} = \frac{\mathrm{a}}{2}\left(\frac{1 - \mathrm{a}^2}{r} + 8\pi r \mathrm{a}^2 \rho\right) , \quad (92)$$

while an equation for the lapse function $\alpha$ follows from the choice of the polar-areal gauge which implies $K_{\theta\theta} = 0$. We find:

$$\partial_r \alpha = \alpha\left(\frac{\mathrm{a}^2 - 1}{2r} + 4\pi r \mathrm{a}^2 S^r_r\right) . \quad (93)$$

In the last two equations, the total energy density $\rho$ and radial stress component $S^r_r$ must include contributions from both the Maxwell and Dirac fields, and are given by:

$$\rho = \rho_M + \rho_D = \frac{1}{8\pi} E_r E^r + \frac{1}{2\pi}\left(\frac{\omega}{\alpha} - q\phi\right)\left(f^2 + g^2\right) , \quad (94)$$

and:

$$\begin{aligned} S^r_r &= S^r_{r\,M} + \mathcal{S}^r_{r\,D} = -\frac{\mathrm{a}^2}{8\pi}\left(E^r\right)^2 + \frac{1}{2\pi \mathrm{a}}\left(f\partial_r g - g\partial_r f\right) \\ &= -\frac{\mathrm{a}}{8\pi}\left(E^r\right)^2 + \rho_D - \frac{1}{\pi}\left(\frac{fg}{r} + \frac{m}{2}\left(f^2 - g^2\right)\right) , \end{aligned} \quad (95)$$

where in the last line we have used equation (88) in order to eliminate the radial derivatives of $f$ and $g$. On the other hand, one can show that:

$$J^r = J^{(M)}_r + \mathcal{J}^{(D)}_r = 0 , \quad (96)$$

as expected for a static spacetime.

As we are considering a spherically symmetric system, there are no magnetic fields and we have the following equations for the scalar potential $\phi$ and the radial component of the electric field $E^r$. First, we can see from equation (21) that $\partial_t \phi = 0$, so that $\phi = \phi(r)$. From equation (22) we then find:

$$\alpha E_r = -\partial_r(\alpha\phi) , \quad (97)$$

or equivalently:

$$\partial_r \vartheta = -\alpha \mathrm{a}^2 E^r , \quad (98)$$

where we have defined $\vartheta := \alpha\phi$. On the other hand, from equation (25) we obtain:

$$\partial_r E^r + E^r\left(\frac{2}{r} + \frac{\partial_r \mathrm{a}}{\mathrm{a}}\right) = 2q\left(f^2 + g^2\right) . \quad (99)$$

Our final system of equations is then:

$$\partial_r \mathrm{a} = \frac{\mathrm{a}}{2}\left(\frac{1-\mathrm{a}^2}{r} + 8\pi r \mathrm{a}^2 \rho\right) , \tag{100}$$

$$\partial_r \alpha = \alpha\left(\frac{\mathrm{a}^2-1}{2r} + 4\pi r \mathrm{a}^2 S_r^r\right) , \tag{101}$$

$$\partial_r f = -f\left(\frac{\partial_r \alpha}{2\alpha} + \frac{1}{r}(1-\mathrm{a})\right) - \mathrm{a}g\left(m + \frac{\omega}{\alpha} - q\phi\right) , \tag{102}$$

$$\partial_r g = -g\left(\frac{\partial_r \alpha}{2\alpha} + \frac{1}{r}(1+\mathrm{a})\right) - \mathrm{a}f\left(m - \frac{\omega}{\alpha} + q\phi\right) , \tag{103}$$

$$\partial_r \vartheta = -\alpha \mathrm{a}^2 E , \tag{104}$$

$$\partial_r E = -E\left(\frac{2}{r} + \frac{\partial_r \mathrm{a}}{\mathrm{a}}\right) + 2q\left(f^2+g^2\right) , \tag{105}$$

where in order to simplify the notation we have taken $E := E^r$, and with:

$$\rho = \frac{\mathrm{a}^2 E^2}{8\pi} + \frac{1}{2\pi\alpha}(\omega - q\vartheta)\left(f^2+g^2\right) , \tag{106}$$

$$S_r^r = -\frac{\mathrm{a}^2 E^2}{8\pi} + \frac{1}{2\pi\alpha}(\omega - q\vartheta)\left(f^2+g^2\right) - \frac{1}{\pi}\left(\frac{fg}{r} + \frac{m}{2}\left(f^2-g^2\right)\right) . \tag{107}$$

## V. BOUNDARY CONDITIONS

In order to solve the above system of equations we have to impose boundary conditions at the origin ($r \to 0$) where the solutions must be regular, and at infinity ($r \to \infty$) where we have to recover Minkowski spacetime. In particular, since we want to describe a compact object, the Maxwell fields have to vanish at infinity, so that for very large $r$ we must have

$$\alpha, \mathrm{a} \to 1 , \qquad \phi, E \to 0 . \tag{108}$$

With the above conditions, the equations for $f$ and $g$ for large $r$ reduce to:

$$\partial_r f \approx -g\,(m+\omega) , \qquad \partial_r g \approx -f\,(m-\omega) . \tag{109}$$

Taking now the second derivative of $f$, and using the equation for $g$, we find:

$$\partial_r^2 f \approx \left(m^2 - \omega^2\right) f , \tag{110}$$

and a similar equation for $g$. Therefore, if we want to find solutions that decay exponentially at infinity we must have $m^2 > \omega^2$. But it is clear that even in that case in general we will also find solutions that grow exponentially. This implies that there will only be purely decaying solutions for very specific values of $\omega$, so that we will need to solve an eigenvalue problem.

Let us now consider the boundary conditions at the origin. As we have already mentioned, the functions $f$ and $g$ behave as $f \simeq f_0 + \mathcal{O}(r^2)$ and $g \simeq g_1 r + \mathcal{O}(r^3)$ near the origin. Then, as $r \to 0$ we will have $f \to f_0$ and $g \to 0$. Also, since spacetime must be locally flat, the radial metric component close to the origin has to behave as $\mathrm{a} \simeq 1 + \mathcal{O}(r^2)$, while the lapse function should behave as $\alpha \simeq \alpha_0 + \mathcal{O}(r^2)$, where $\alpha_0$ is some constant. On the other hand, from equation (105) we can see that near the origin we have:

$$\partial_r E = -2\,\frac{E}{r} + 2q f_0^2 , \tag{111}$$

which be rewritten as:

$$\frac{d}{dr}(E r^2) = \lambda^2 r^2 , \tag{112}$$

with $\lambda^2 = 2qf_0^2$. Integrating we find:

$$E = \frac{\lambda^2 r}{3} + \frac{c}{r^2} \; , \tag{113}$$

with $c$ an integration constant. Demanding regularity now requires $c = 0$, so that we finally find that near the origin we must have:

$$E \approx \frac{\lambda^2 r}{3} \; . \tag{114}$$

Since clearly $E$ vanishes for $r = 0$, equation (104) now implies that $\vartheta \simeq \vartheta_0 + \mathcal{O}(r^2)$, with $\vartheta_0$ constant.

We then have the following set of conditions at $r = 0$:

$$\mathrm{a}(0) = 1 \; , \quad \alpha(0) = \alpha_0 \; , \quad f(0) = f_0 \; , \quad g(0) = 0 \; , \quad \vartheta(0) = \vartheta_0 \; , \quad E(0) = 0 \; , \tag{115}$$

together with:

$$\partial_r \mathrm{a}(0) = 0 \; , \quad \partial_r \alpha(0) = 0 \; , \quad \partial_r f(0) = 0 \; , \quad \partial_r g(0) = \left(\frac{\omega}{\alpha} - q\phi - m\right)\frac{f_0}{3} \; , \quad \partial_r \vartheta(0) = 0 \; , \quad \partial_r E = \frac{2qf_0^2}{3} \; . \tag{116}$$

We can also see that the system of equations is invariant under the following rescaling:

$$\alpha \to C\alpha \; , \qquad \vartheta \to C\vartheta \; , \qquad \omega \to C\omega \; , \tag{117}$$

with $C$ an arbitrary constant. This implies that we can solve the system by initially taking $\alpha_0 = 1$, and after the numerical integration we can rescale $\alpha$, the frequency $\omega$, and $\vartheta = \alpha\phi$ as:

$$\alpha \to \alpha/\alpha_\infty \; , \qquad \vartheta \to \vartheta/\alpha_\infty \; , \qquad \omega \to \omega/\alpha_\infty \; , \tag{118}$$

where $\alpha_\infty$ is the asymptotic value of the lapse function obtained from the original solution.

On the other hand, for the scalar potential we can initially choose $\vartheta_0 = 0$. In order for our final solution to be compatible with the condition that $\vartheta(r \to \infty) = 0$, after the numerical integration we can make a gauge transformation of the following form (see reference [5]):

$$\vartheta \to \vartheta - \partial_t \theta \; , \qquad \omega \to \omega + q\partial_t \theta \; , \tag{119}$$

with $\theta = \vartheta_\infty t$, and $\vartheta_\infty$ the asymptotic value of $\vartheta$ obtained after the rescaling given by (118).

In summary, we solve the system of equations (100)-(105) using the boundary conditions given by (108) at infinity and (115)-(116) at the origin, taking initially $\alpha_0 = 1$, $\vartheta = 0$ at $r = 0$. And after the numerical integration we first rescale $(\alpha, \vartheta, \omega)$ using (118), and finally perform the gauge transformation (119).

## VI. TOTAL MASS, CHARGE AND BINDING ENERGY.

In this section we introduce the global quantities that characterize the stationary self-gravitating solutions, namely the total mass, total electric charge, and binding energy. Once we have solved the system of equations (100)-(105) we can compute these global quantities numerically.

The total mass $M$ is given by the integral of the total energy density over a flat volume element:

$$M = 4\pi \int_0^\infty \rho r^2 dr \; , \tag{120}$$

with $\rho$ given by equation (106). This expression can be derived directly from the Hamiltonian constraint when one uses the areal radius. The use of the flat volume element in the above expression comes out naturally from the Hamiltonian constraint when working in this gauge, and is related to the fact that the total mass $M$ is not simply the integral of the energy density over the physical (curved) space, but must include the (negative) contribution from the gravitational potential energy.

On the other hand, the total electric charge $Q$ is correctly given by the integral of the charge density over the physical volume element, and correspond to the conserved Noether charge:

$$Q = 4\pi \int_0^\infty \rho_q^{\mathrm{Tot}} \mathrm{a} r^2 dr \; . \tag{121}$$

And using the fact that $\rho_q^{\rm Tot} = q(f^2+g^2)/2\pi$ we finally find:

$$Q = 2q \int_0^\infty (f^2+g^2) a r^2 dr \; . \tag{122}$$

Having found the total charge, one can define the total number of particles simply by $N = Q/q$. Notice that in the case of fermions this number is in fact not easy to interpret since the Pauli exclusion principle should imply that we can only have one fermion in each quantum state, and since here we are using two spinors in order to guarantee spherical symmetry, we should have only two fermions. However, we must stress the fact that here we are considering the Dirac field as a purely "classical" field, and one can interpret $N$ as an effective particle number.

Notice that for Dirac stars the spinor fields $(f,g)$ decay exponentially, so that at large distances only the electromagnetic field remains and the spacetime should approach rapidly the Reissner–Nordström solution. In that case we will have:

$$a^2(r) \to \left( 1 - \frac{2M}{r} + \frac{Q^2}{r^2} \right)^{-1} \; . \tag{123}$$

We can then define the Reissner–Nordström mass as:

$$M_{\rm RN} = \lim_{r\to\infty} \left[ \frac{r}{2} \left( 1 + \frac{Q^2}{r^2} - \frac{1}{a^2} \right) \right] \; . \tag{124}$$

At infinity both definitions for the mass, (120) and (124), should clearly coincide. However the mass integral (120) converges quite slowly since the electric field is long range and contributes to the total energy density. On the other hand, the Reissner–Nordström mass converges to the total mass very fast since the spinor fields decay exponentially. Because of this, from here on we will use the Reissner–Nordström mass (124) to compute the quantities of interest (but we have indeed verified that for very large values of $r$ both versions of the mass approach each other).

Finally, we define the binding energy $E_B$ of a given configuration in the usual way:

$$E_B = M_{\rm RN} - mN = M_{\rm RN} - m \left( \frac{Q}{q} \right) \; . \tag{125}$$

This quantity measures the difference between total mass of the the self-gravitating configuration, which includes contributions from kinetic, potential and rest mass energy, and the rest mass of a collection of $N$ free particles each of mass $m$. If we have $E_B < 0$ the solution is gravitationally bound, whereas $E_B > 0$ indicates that the configuration, though time independent, is not gravitationally bound and can therefore be expected to be unstable.

Notice that the binding energy $E_B$ can also be written as:

$$E_B = M_{\rm RN} \left[ 1 - \left( \frac{m}{q} \right) \left( \frac{Q}{M_{\rm RN}} \right) \right] \; . \tag{126}$$

From this equation we can see that if we have $Q/M_{RN} > q/m$ we will obtain gravitationally bound configurations, while for $Q/M_{RN} < q/m$ the solutions will be unbound.

## VII. NUMERICAL RESULTS

We solve the system of ordinary differential equations (100)-(105) numerically using a fourth order Runge–Kutta method, with the boundary conditions (115)-(116), taking initially $\alpha_0 = 1$ and $\vartheta_0 = 0$ at the origin and later applying the rescaling and gauge transformation described in Sec. V above. Notice that there are three free parameters in our system of equations: the value of the spinor amplitude $f(r)$ at the origin $f_0$, plus the values of the charge $q$ and mass $m$ parameters of the Dirac field. In the same way as in [5], in order to solve the system we always take $m = 1$ (but notice that the system of equations can be rescaled to arbitrary values of $m$), fix a value of the charge $q$, and vary the value of $f_0$.

For each value of $q$ and $f_0 \in [0.01, 1]$ we use a shooting algorithm in order to determine the value of the frequency $\omega$ that corresponds to exponential decay far away. In principle, for a given value of $f_0$ there is an infinite spectrum of possible values of $\omega$ corresponding to solutions with increasing number of nodes, but here we concentrate on the solution with no nodes corresponding to the lower energy solution (the ground state).

For all the solutions reported here we have used a grid spacing of $\Delta r = 0.01$. Also, for all values of the charge parameter such that $q \neq 1$, the outer boundary is located at $r = 300$. However, for the special case with $q = 1$, the outer boundary is fixed at $r = 300$ for $f_0 \in [0.05, 1]$, at $r = 400$ for $f_0 = (0.03, 0.04)$, at $r = 500$ for $f_0 = 0.02$, and at $r = 1000$ for $f_0 = 0.01$. The reason for this is that for the special case of $q = 1$, as $f_0 \to 0$ both the radius of the star and its total mass $M$ keep increasing, requiring the computational boundary to be placed farther away in order to properly characterize the solution.

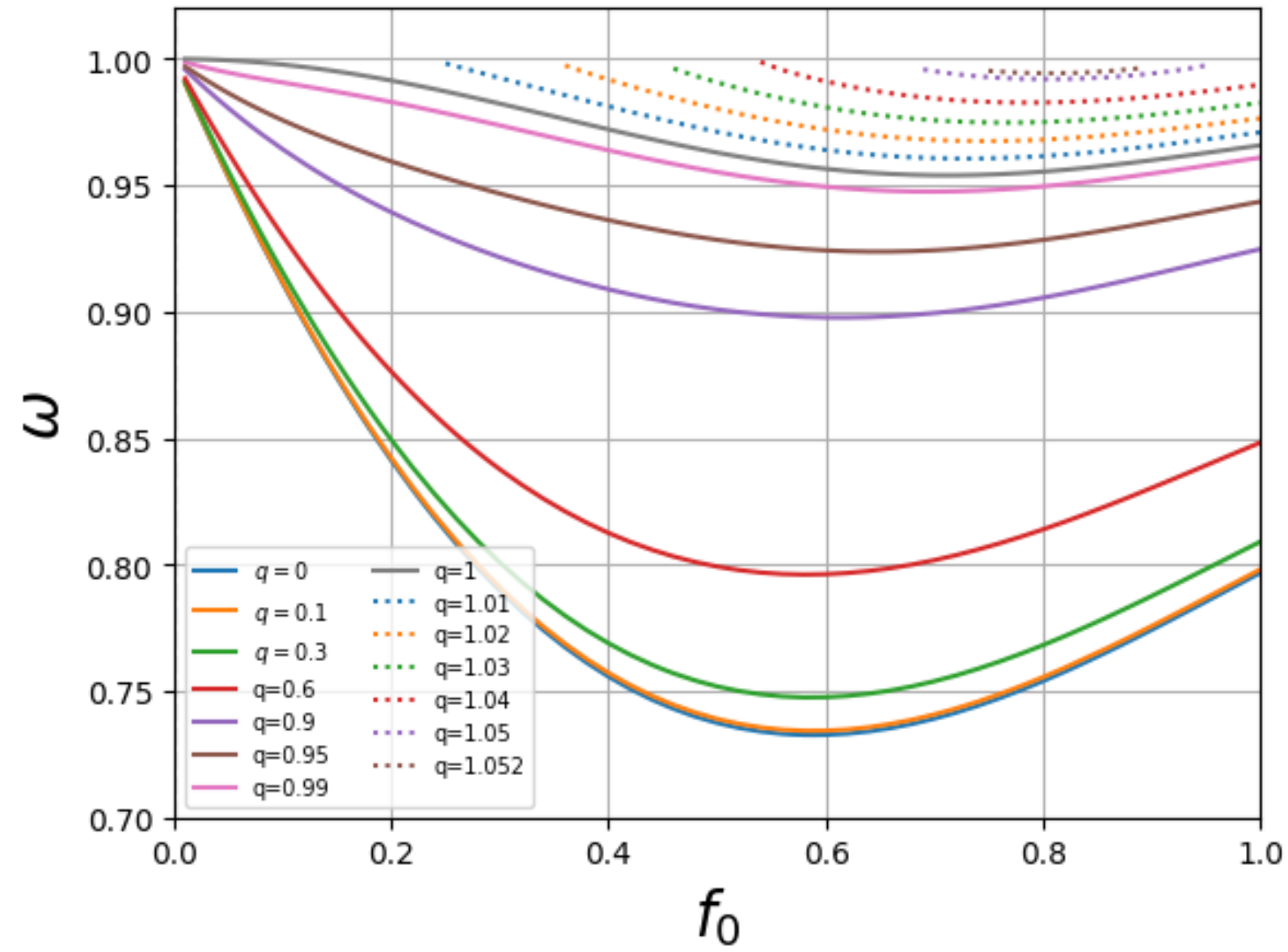

**Figure 1:** Frequency $\omega$ as a function of $f_0$ for different values of $q$. For subcritical solutions with $q \leq 1$ the lowest value of the frequency is $\omega = 0.7327$ for $q = 0$ and $f_0 = 0.59$. We do find super-critical solutions for $1 < q \leq 1.052$, but only for a limited range of values of $f_0$.

### A. Frequency

In Figure 1 we plot the value of the frequency $\omega$ as a function of $f_0$, for different values of $q$. As can be seen from the plot, we find that for $q \leq 1$ the value of the frequency for small $f_0$ approaches 1. But as we increase $f_0$ the frequency decreases until it reaches a minimum, and then increases again. Also, as the value of the charge $q$ increases this minimum frequency becomes larger. On the other hand, for charges such that $q > 1$ we find solutions only for a small range of values of $f_0$, and this range decreases as the value of $q$ increases further. Furthermore, we can only find solutions for $1 < q \leq 1.052$. It is important to note here that, at least at the Newtonian level, $q > 1$ corresponds to a regime in which the Coulomb repulsion overcomes gravity, and thus no stationary solutions should be expected. However, it has recently been shown that both in the case of charged boson stars [16] and charged Proca stars [5], such super-critical solutions also exist in the general relativistic case, but only for a limited range of values of $q$ and $f_0$. Nevertheless, even though solutions are present in this super-critical regime, their binding energy is always positive (see below), and thus they do not represent gravitationally bound configurations.

### B. Effective radius

As we are looking for solutions that decay exponentially at large distances, charged Dirac stars do not possess a well-defined radius or a sharp surface that bounds the matter distribution. Therefore in this work, and following previous studies, we adopt the concept of an effective radius defined as the radius that encloses 99% of the total charge $Q$ of the star, which we denote as $R_{99}$. In contrast to the cases with no charge $q = 0$, we prefer not to use the radius that contains 99% of the total mass, since for $q \neq 0$ the electric field contributes to the total mass and it decays very slowly.

Figure 2 shows a plot of the effective radius $R_{99}$ as a function of $f_0$ for the different values of $q$. Notice that for a fixed value of $q$, small values of $f_0$ correspond to stars with very larger radius, with the radius becoming smaller as we increase $f_0$. Notice also that as we increase the value of $q$, the stars always have larger radius for the same value of $f_0$, which is to be expected since the Coulomb repulsion works against gravity. In particular, the larger values of $R_{99}$ are obtained for the case of $q = 1$ and very small $f_0$.

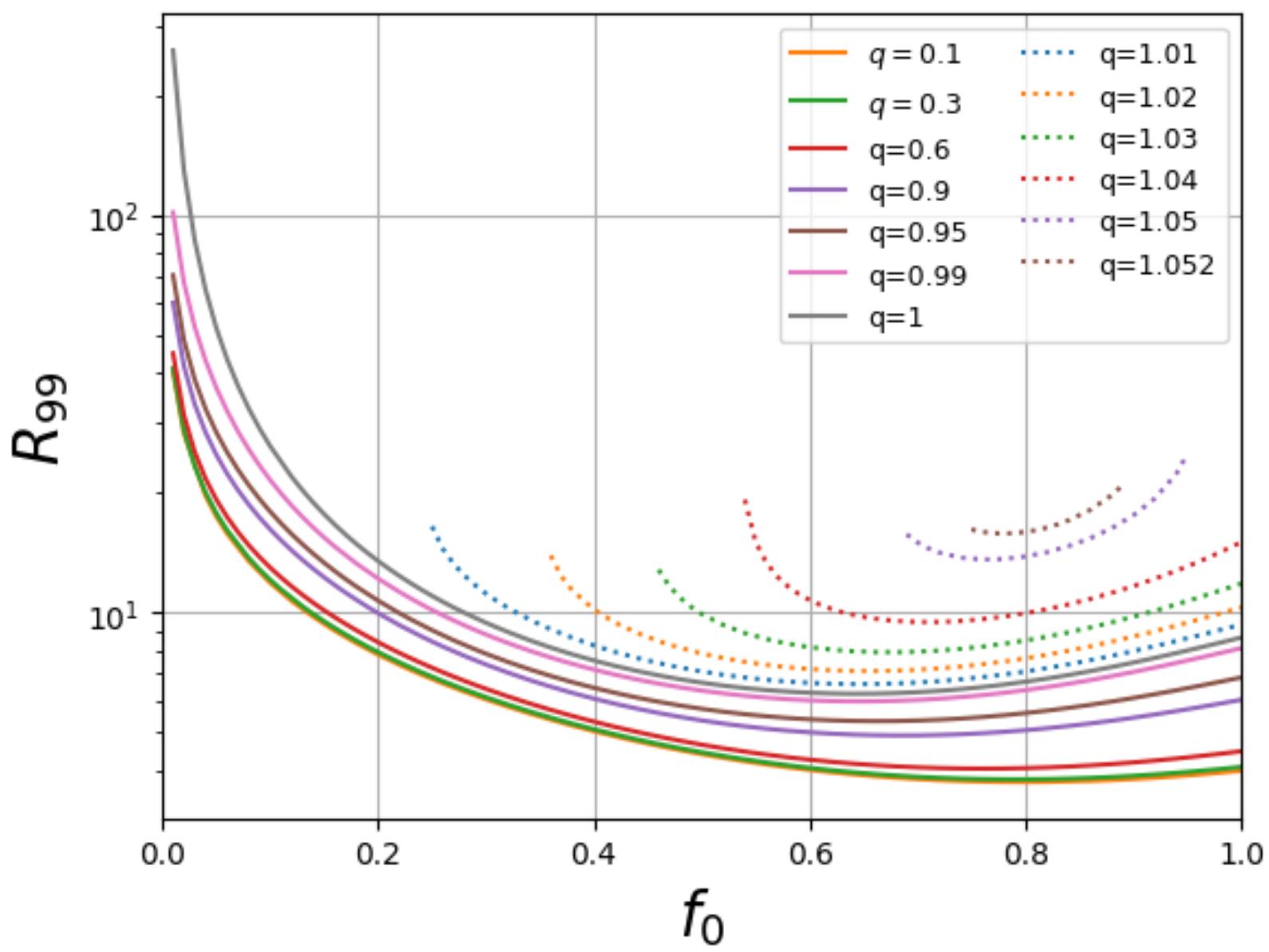

**Figure 2:** Effective radius $R_{99}$ as a function of $f_0$ for different values of $q$. For $q \leq 1$, the radius becomes larger as $f_0 \to 0$, but decreases with increasing $f_0$ until reaching a minimum value, and then increases again as $f_0$ becomes even larger. Also, as the charge $q$ increases the values of $R_{99}$ increase for the same value of $f_0$.

### C. Total mass and compactness

In Figure 3 we plot the variation of the total Reissner–Nordström mass $M_{RN}$ with respect to the frequency $\omega$ for different values of $q$. For $q \leq 0.95$, the plot of the mass $M_{RN}$ with respect to $\omega$ describes a well-defined spiral as we change the value of $\omega$. Also, as the charge $q$ increases the mass generally increases. The curves corresponding to $q = 0$ and $q = 0.1$ are almost indistinguishable, indicating that small charges have a negligible effect on the mass-frequency relation. For the case $q = 0.99$ the spiral is shrinks considerably and is shifted toward larger values of $\omega$.

For the special $q = 1$ the spiral structure in fact disappears and the mass increases steeply as $\omega \to 1$. This can also be seen in Figure 4 where we plot the total mass with respect to $f_0$. Notice how for $q < 1$ there is a clear maximum of the mass the becomes larger and moves to smaller values of $f_0$ as we increase the value of $q$. However, for $q = 1$ there is no clear maximum. In fact, we have found that for $q = 1$ the total mass seems to increase without bound as we decrease the value of $f_0$ even further, while the effective radius also increases. This might seem counter-intuitive since the limit $f_0 = 0$ corresponds to having no star at all. However, it would seem that in this case the limit is not regular.

For $q > 1$ the solutions no longer develop the full spiral structure observed for $q < 1$. Instead, they exhibit only incomplete spiral branches that become progressively shorter as the charge increases further. Near the upper limit of the allowed charge, $q \simeq 1.052$, the solutions are confined to a small loop in the $(\omega, M_{RN})$ plane (Figure 3b).

The compactness of our configurations is defined as $C = M_{RN}/R_{99}$, and measures how concentrated is the mass distribution within the radius $R_{99}$. The value of this quantity depends on the value of the charge $q$ and on $f_0$. This can be seen in Figure 5, where we plot the compactness $C$ as a function of $f_0$. For $q < 1$, we find that $C$ approaches 0 for small $f_0$, becomes larger as $f_0$ increases until it reaches a maximum value, and then decreases again as $f_0 \to 1$. We also plot the compactness for the uncharged case $q = 0$, but since in this case there is no charge, its compactness $C$ is defined simply as $C := M/R_{99}$ where $M$ is the integrated mass and $R_{99}$ is the radius that contains 99% of this mass. We can see from the left panel of Figure 5 that, for $f_0 \in [0.8, 1]$ and $q = 0$, the values of the compactness are in fact greater than those of the solutions with $q > 0$. Furthermore for $q < 1$, the maximum values for the compactness increase as the charge increases.

As before, the case $q = 1$ behaves differently, with the compactness having a maximum value as $f_0 \to 0$, reaching a limit of $C \sim 0.302$. We can compare these values with the compactness of a black hole $C = 0.5$, and with that of a neutron star $C \sim 0.3$ [17].

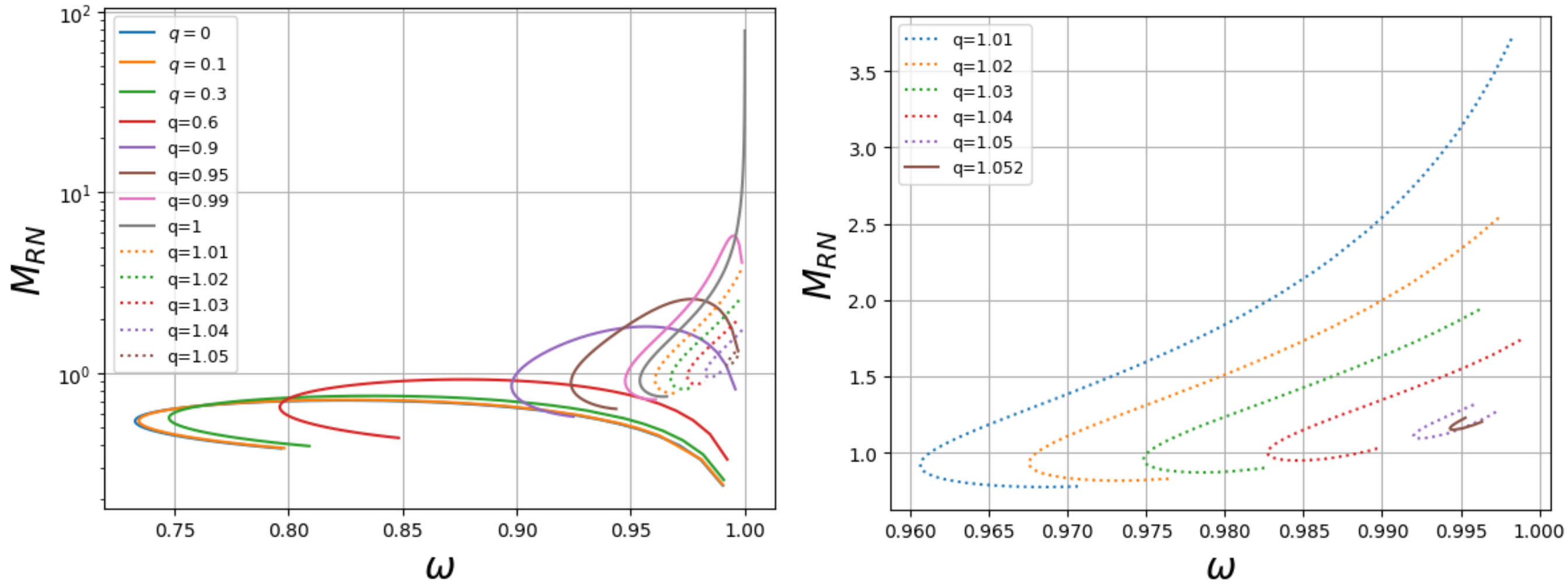

**Figure 3:** Left panel: Total mass $M_{RN}$ with respect to $\omega$ for different values of q. For $q \leq 0.95$ the solutions describe a clear spiral, while for $q = 0.99$ the spiral shrinks and moves to larger values of $\omega$. For $q = 1$ the spiral structure is lost, and the mass appears to grow without bound as $\omega \to 1$ and $f_0 \to 0$. The curves for $q = 0$ and $q = 0.1$ are nearly indistinguishable. Right panel: We show only the super-critical cases with $q > 1$ for clarity.

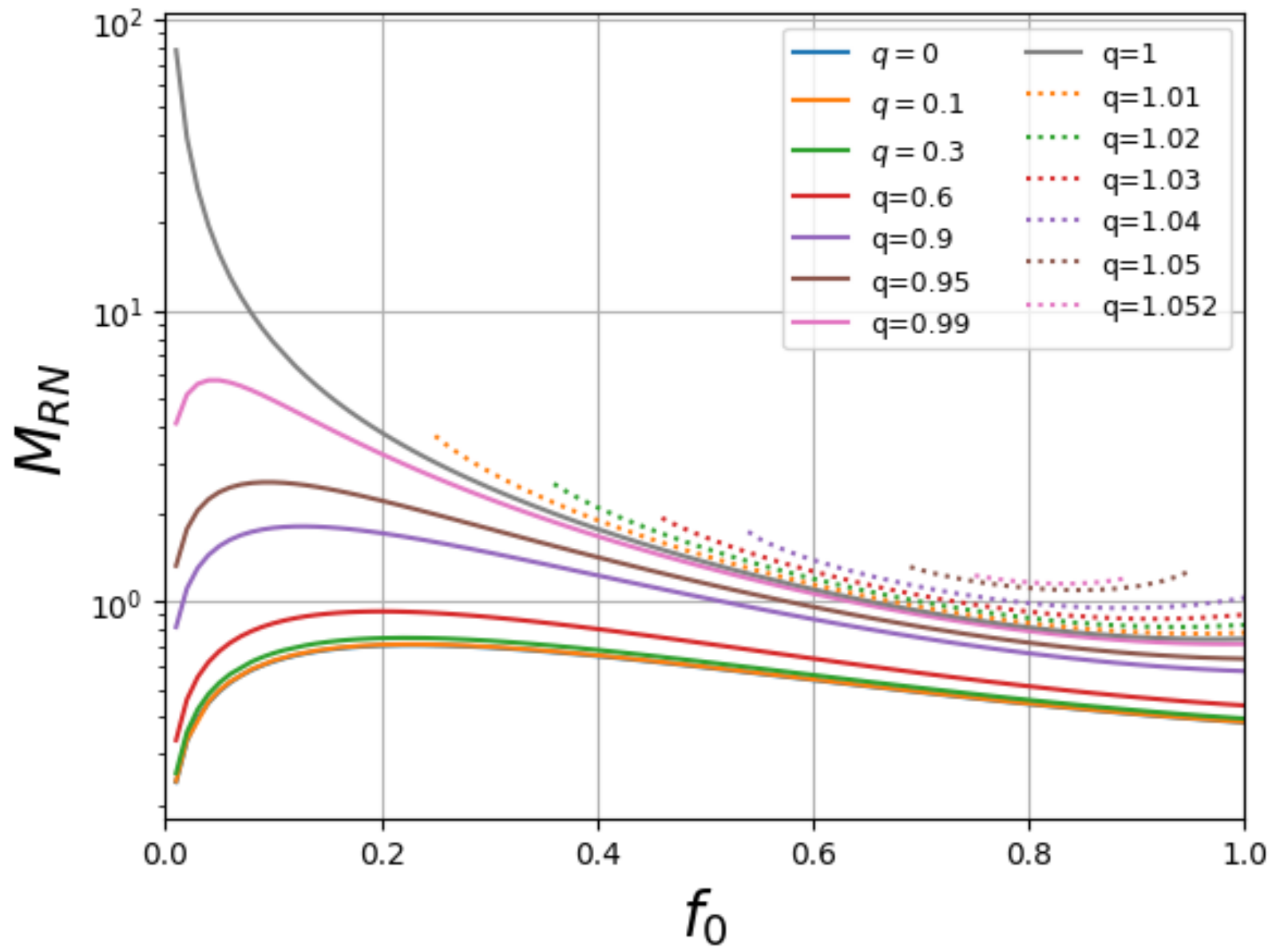

**Figure 4:** Total mass $M_{RN}$ with respect to $f_0$ for different values of q. For the cases with $q < 1$ the maximum value of the mass increases and moves to smaller values of $f_0$. On the other hand, for the special case with $q = 1$ the mass seems to increase without bound as $f_0$ approaches 0.

### D. Binding energy and total charge

In Figure 6 we plot the binding energy $E_B$ defined in equation (125) as a function of $f_0$ for the different values of $q$ (upper-left panel). We can see from this figure that for $q < 1$ there exist intervals of $f_0$ for which $E_B < 0$, corresponding to gravitationally bound configurations. These intervals become progressively smaller as the charge increases. In Table I we show the values of $(f_0, \omega, M_{RN}, Q, N, R_{99}, C)$ corresponding to the minimum value of the binding energy $E_B$ for each value of the charge parameter $q$. This minimum of the binding energy always corresponds with the configuration with maximum mass in each family (see below). We also show in Table II the values of $(f_0, \omega, M_{RN}, Q, N, R_{99}, C)$ for which the binding energy becomes zero. On the other hand, for $q \geq 1$ the binding energy is always positive, which indicates that for these values of the charge there are not bound states. Furthermore, the binding energy clearly increases with $q$.

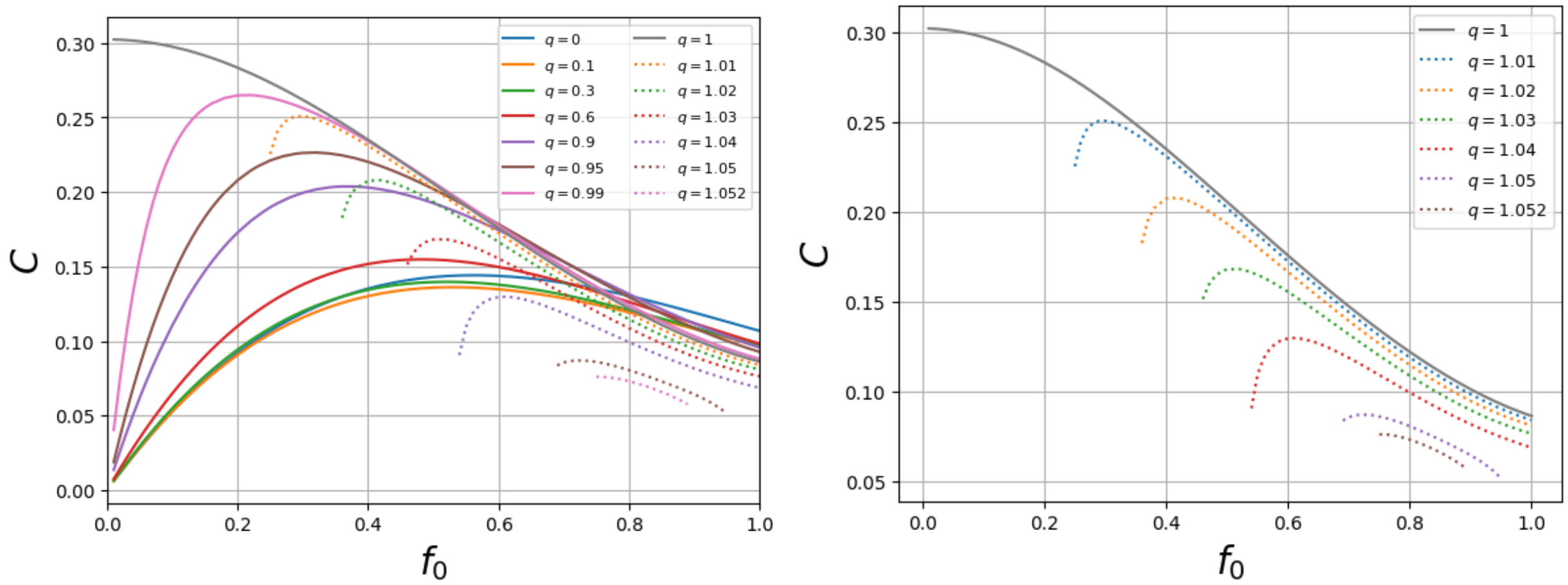

**Figure 5:** Left panel: Compactness $C$ with respect to $f_0$ for different values of $q$. For $q < 1$ the compactness has a clear maximum that becomes larger and moves to the left as $q$ increases. On the other hand, for $q = 1$ the maximum compactness is reached as $f_0 \to 0$, with a value of $C \sim 0.302$. Right panel: We show only the cases with $q \geq 1$ for clarity.

| $q$ | $f_0$ | $\omega$ | Minimum $E_B$ | $M_{RN}$ | $Q$ | $N$ | $R_{99}$ | $C$ |
|---|---|---|---|---|---|---|---|---|
| 0 | 0.23 | 0.8240 | $-2.795 \times 10^{-2}$ | 0.7089 | 0 | 0.7369 | 6.999 | 0.1013 |
| 0.1 | 0.23 | 0.8250 | $-2.788 \times 10^{-2}$ | 0.7133 | 0.0741 | 0.7411 | 7.166 | 0.0995 |
| 0.3 | 0.22 | 0.8383 | $-2.732 \times 10^{-2}$ | 0.7507 | 0.2334 | 0.7781 | 7.474 | 0.1005 |
| 0.6 | 0.20 | 0.8764 | $-2.484 \times 10^{-2}$ | 0.9252 | 0.5700 | 0.9500 | 8.411 | 0.1100 |
| 0.90 | 0.13 | 0.9559 | $-1.602 \times 10^{-2}$ | 1.8135 | 1.6466 | 1.8295 | 13.546 | 0.1339 |
| 0.95 | 0.09 | 0.9777 | $-1.195 \times 10^{-2}$ | 2.5716 | 2.4544 | 2.5836 | 17.934 | 0.1332 |
| 0.99 | 0.04 | 0.9955 | $-5.567 \times 10^{-3}$ | 5.7646 | 5.7125 | 5.7702 | 43.130 | 0.1337 |

**Table I:** Values of $(f_0, \omega, M_{RN}, Q$,N$, R_{99}, C)$ corresponding to the minimum value of the binding energy $E_B$ (maximum value of the total mass) for different values of $q$.

| $q$ | $f_0$ | $\omega$ | $M_{RN}$ | $Q$ | $N$ | $R_{99}$ | $C$ |
|---|---|---|---|---|---|---|---|
| 0 | 0.46 | 0.7431 | 0.6171 | 0 | 0.6171 | 4.383 | 0.1408 |
| 0.1 | 0.46 | 0.7446 | 0.6199 | 0.6201 | 0.0620 | 4.611 | 0.1345 |
| 0.3 | 0.45 | 0.7588 | 0.6516 | 0.1953 | 0.6512 | 4.722 | 0.1380 |
| 0.6 | 0.4 | 0.8126 | 0.8052 | 0.4835 | 0.8059 | 5.307 | 0.1517 |
| 0.90 | 0.26 | 0.9280 | 1.5767 | 1.4194 | 1.5771 | 8.202 | 0.1922 |
| 0.95 | 0.20 | 0.9595 | 2.2245 | 2.1129 | 2.2242 | 10.700 | 0.2079 |
| 0.99 | 0.09 | 0.9914 | 5.1231 | 5.0727 | 5.1240 | 23.452 | 0.2184 |

**Table II:** Values of $(f_0$, $\omega$, $Q$,$M_{RN}$,$Q$,$N$, $R_{99}$, $C)$ for which binding energy vanishes, $E_B = 0$, for the different values of $q$.

The upper-right panel of Figure 6 shows the binding energy $E_B$ as a function of $\omega$. The lower-left panel shows the total mass $M_{RN}$ with respect to the binding energy for $q < 1$. One can clearly see that the maximum mass $M_{RN}$ is always attained when the binding energy $E_B$ has a minimum. Finally, the lower-right panel shows the variation of the compactness $C$ with respect to $E_B$ for different values of $q$. Both bound and unbound configurations are present, and for all values of the charge there exists values of $C$ comparable to those of neutron stars.

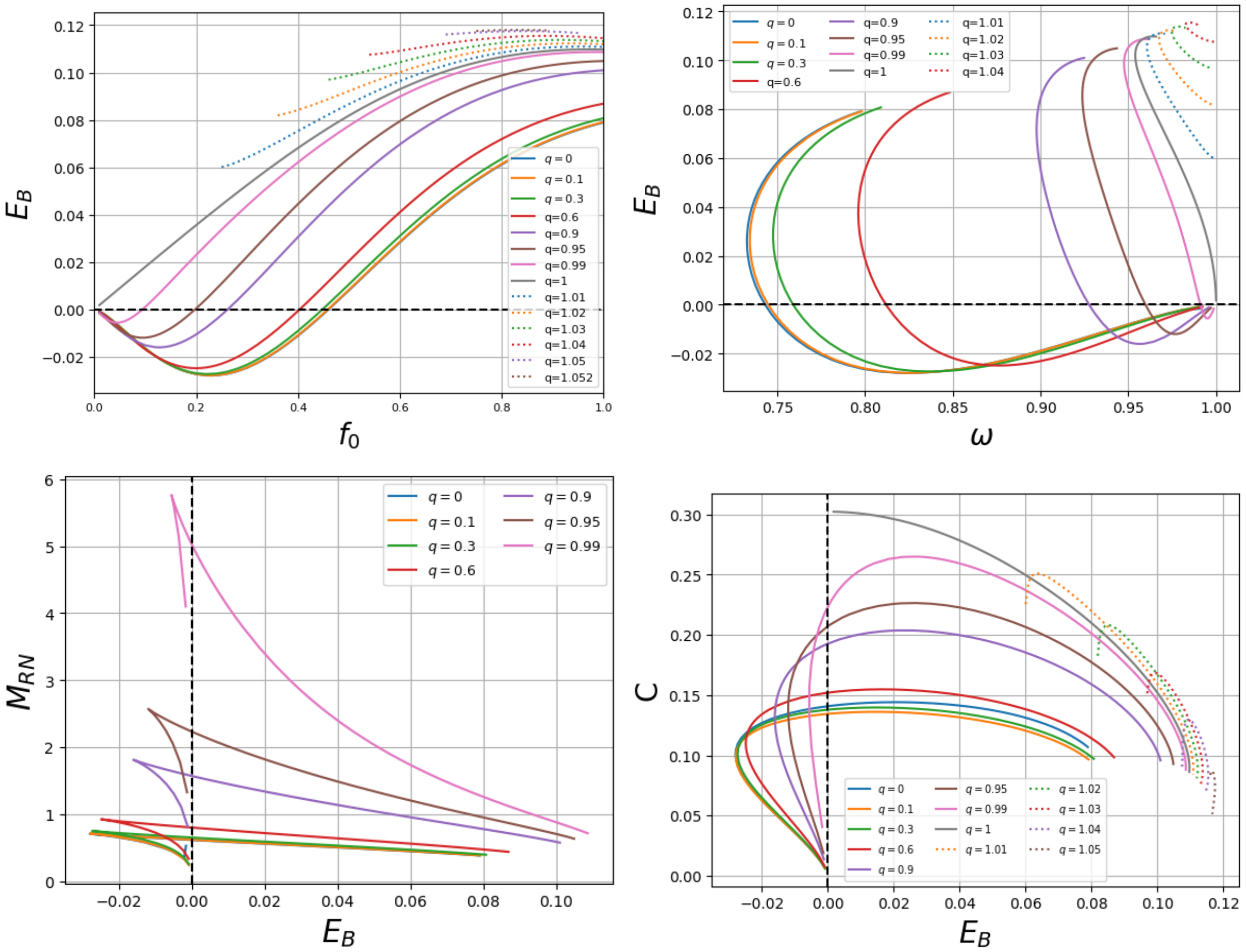

**Figure 6:** Upper left panel: Binding energy $E_B$ as a function of $f_0$ for the different values of $q$. For $q < 1$ there exist intervals of $f_0$ for which $E_B < 0$, and the range of these intervals decreases as the charge increases. However, for $q \geq 1$ we always have $E_B > 0$. Upper right panel: Variation of $E_B$ with respect to $\omega$. Lower left panel: Variation $M_{RN}$ with respect to binding energy $E_B$. When the binding energy is a minimum the mass $M_{RN}$ is a maximum. Lower right panel: Compactness $C_Q$ as a function of the binding energy $E_B$. For $q < 1$ there exists stationary gravitationally bound solutions with compactness $C \sim 0.1 - 0.2$ comparable to those of neutron stars. For $q \geq 1$, the compactness reaches values in the range $C \sim 0.05 - 0.302$; however, these configurations are not gravitationally bound.

Let us now consider the total charge of our configurations. In Section VI we mentioned that in order to obtain bound solutions we must have $Q/M_{RN} > q/m$. Therefore, Since we have fixed the mass parameter to $m = 1$, we will only have bound solutions when $Q/M_{RN} > q$ is satisfied. In Figure 7 we plot $Q/M_{RN}$ as a function of $f_0$ for the different values of $q$. Notice that all cases we find $Q/M_{RN} < 1$, indicating that the total charge $Q$ is always smaller than the total mass. This explains why we can find some super-critical solutions: even if for those solutions we have $q > m$, we always find $Q < M_{RN}$, so that the total Coulomb repulsive force is smaller than gravity. In particular, we find $Q \sim M_{RN}$ when $q = 1$ and $f_0 \to 0$.

## VIII. SOME PARTICULAR CONFIGURATIONS

In this section we will present some particular solutions for the variables of our system: $f(r)$, $g(r)$, $\alpha(r)$, $a(r)$, $\phi(r)$, $E(r)$. For this purpose we choose four values of the charge parameter, $q = (0.60, 0.90, 1, 1.01)$, and four central value for $f(r)$, $f_0 = (0.06, 0.15, 0.25, 0.60)$.

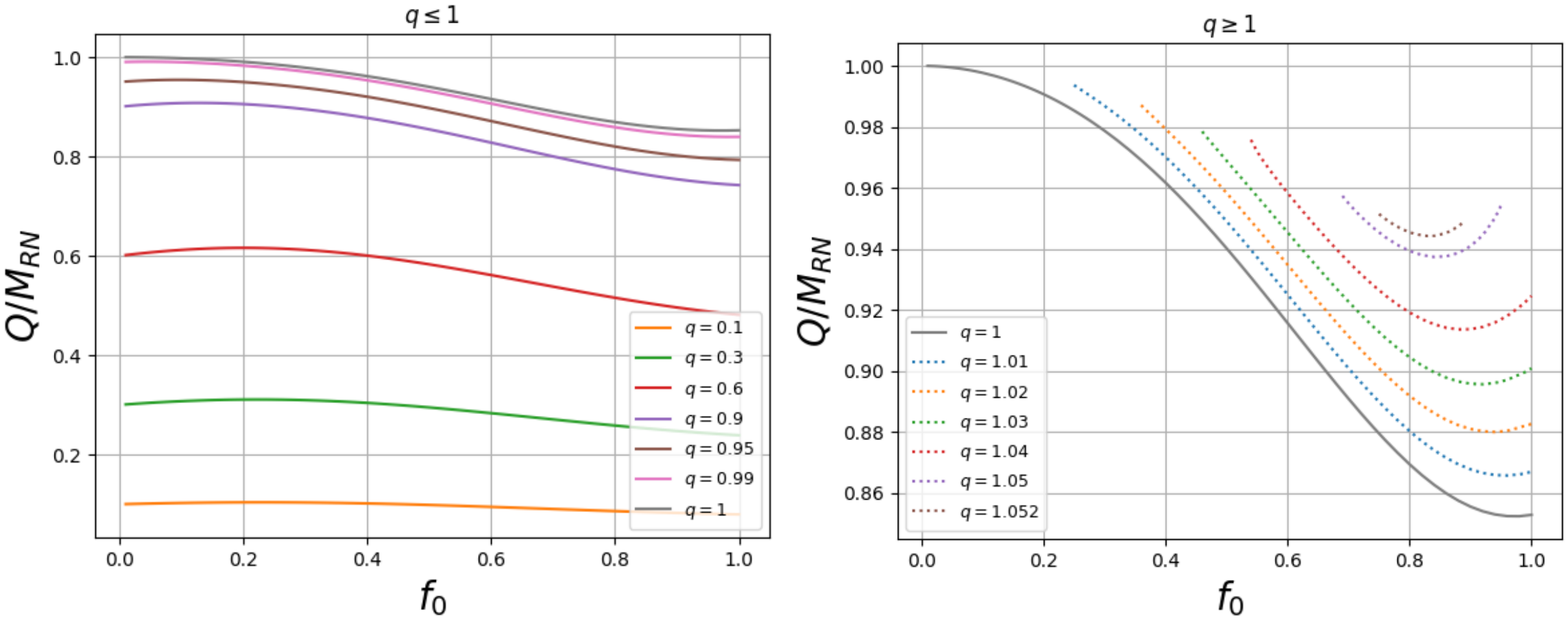

**Figure 7:** Left panel: $Q/M_{RN}$ as a function of $f_0$ for $q \leq 1$, we can see that we always have $Q/M_{RN} < 1$. Right panel: $Q/M_{RN}$ as a function of $f_0$ for $q \geq 1$. Although for these cases we have $q \geq m = 1$, we always find $Q \leq M_{RN}$.

According to equation (72), the spinor solutions are given by:

$$\psi_\pm = \frac{e^{i(\pm\varphi/2-\omega t)}}{\sqrt{4\pi}} \begin{pmatrix} f(r) y_\pm(\theta) \\ \pm i f(r)\, y_\mp(\theta) \\ i g(r)\, y_\pm(\theta) \\ \pm g(r)\, y_\mp(\theta) \end{pmatrix} . \tag{127}$$

The radial profiles of the spinor amplitudes $f(r)$ and $g(r)$ for each of our configurations are plotted in Figure 8. From the figure, we can clearly see that $f(r)$ is a monotonically decreasing function with maximum value $f_0$ at the origin, and which approaches zero exponentially as $r \to \infty$. In contrast, $g(r)$ vanishes at the origin, increases to a maximum and then decreases, approaching zero exponentially at large radii. As $f_0$ decreases, both spinor amplitudes become more extended, indicating that smaller values of $f_0$ correspond to Dirac stars with larger radii. It is important to notice that varying the charge over the range considered produces only minor changes in the radial profiles of both $f(r)$ and $g(r)$, their shape depends primarily on the central amplitude $f_0$.

In Figure 9 we show the lapse function $\alpha(r)$ (left panel) and radial metric function $a(r)$ (right panel) for the different configurations. One can see that, in general, as the charge increases the central value of the lapse $\alpha(0)$ becomes smaller, but in all cases $\alpha \to 1$ for large $r$ and $\alpha'(0) = 0$. Also, we can see that $a(r)$ has a bell-shaped profile, with a maximum that increases as both $q$ and $f_0$ become larger. In all cases we have $a(0) = 1$, $a'(0) = 0$, and $a \to 1$ far away.

Figure 10 shows the electric potential $\phi(r)$ and electric field $E(r)$ for our configurations. We verify that $\phi, E \to 0$ for large $r$, and $E(0) = 0$ with $\partial_r E(0) = 2qf_0^2/3$. As expected, the electric field and potential become stronger for larger values of the charge $q$, and also increase with $f_0$.

Finally, in Figure 11 we plot the energy density $\rho(r)$, which includes contributions from both the Maxwell and Dirac fields as given by equation (94). The total energy density is more concentrated near the origin, and the density profile depends primarily on the central amplitude $f_0$.

## IX. CONCLUSIONS

In this work we solved the Einstein-Dirac-Maxwell (EDM) system within the 3+1 formalism of general relativity in order to construct families of stationary, charged, and spherically symmetric self-gravitating configurations. To this end we considered a spherically symmetric spacetime and two spinor fields minimally coupled to the Maxwell field. We obtained a set of differential equations for the Dirac amplitudes $f(r)$, $g(r)$, the electric field $E(r)$, the electric potential $\phi(r)$, the lapse function $\alpha(r)$ and the radial metric function $a(r)$, that depend only on 3 parameters: the charge parameter $q$, the mass of the Dirac field $m$, and the amplitude of the Dirac field at the origin $f_0$. This system is then solved numerically by fixing the mass to $m = 1$ and varying the charge $q$ and $f_0$, with the boundary conditions present in Section V.

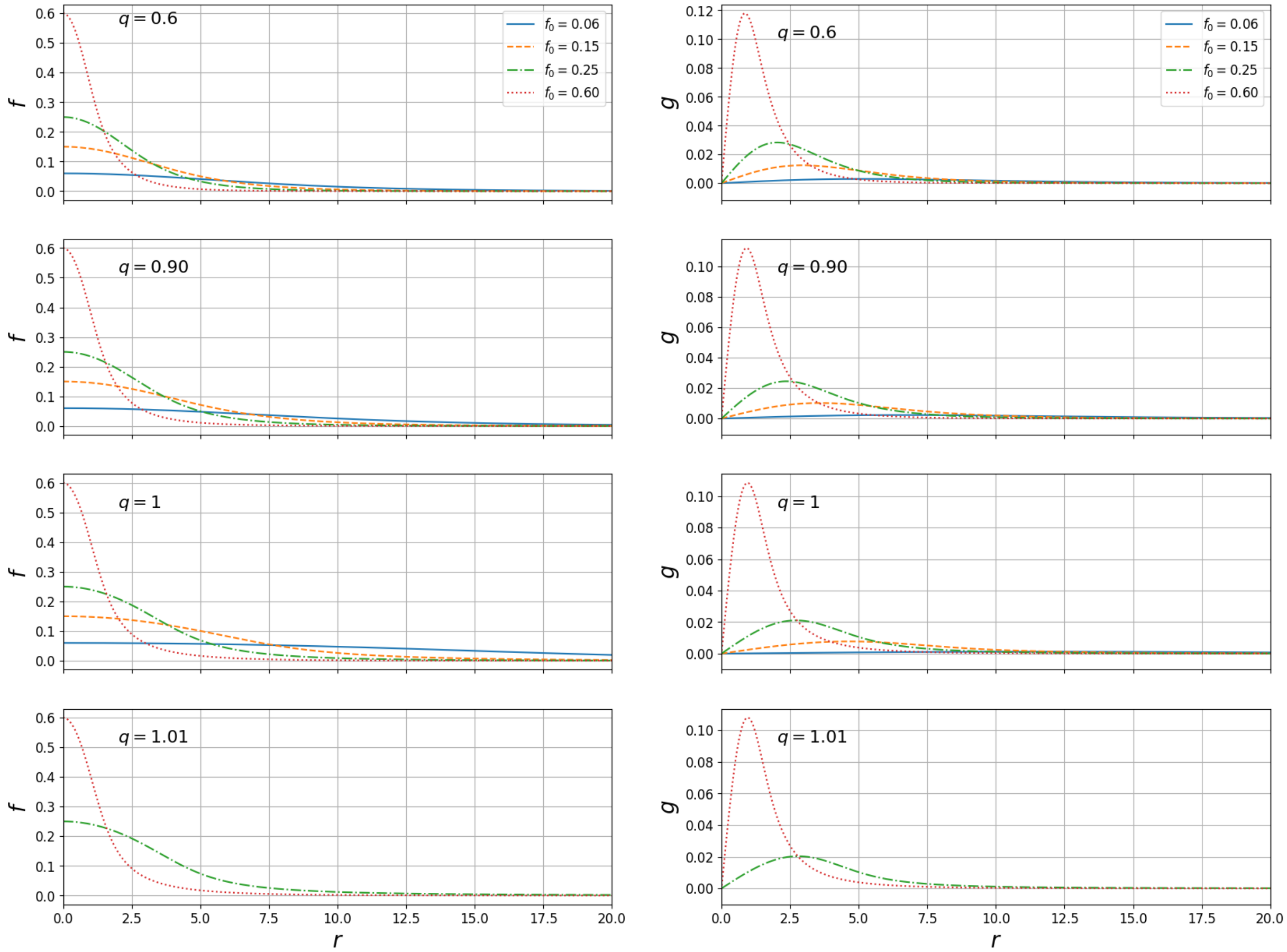

**Figure 8:** Spinor amplitudes $f(r)$ and $g(r)$ with for the different values of $q$ and $f_0$.

We found that for $0 \leq q \leq 1.052$, and $f_0 \in [0.01, 1]$, the Dirac field admits stationary solutions of the form given by equation (127), with the corresponding values of $\omega$ shown in Figure 1. For $1 < q \leq 1.052$, stationary solutions exist only within a narrow range of $f_0$. Additionally, we found that gravitationally bound solutions, characterized by $E_B < 0$, occur only for charges $q < 1$ and within a certain range of values of $f_0$. Moreover, the interval of $f_0$ corresponding to gravitationally bound configurations ($E_B < 0$) decreases as the charge increases, implying that the number of gravitationally bound configurations is progressively reduced. These results are consistent with those previously reported in [18]. As an important result, we find that there are no gravitationally bound configurations for $q \geq 1$.

Furthermore, we defined and computed some interesting properties of these stationary solutions: the total mass $M_{RN}$, the effective radius $R_{99}$, the total charge $Q$, and the compactness $C$, and the binding energy $E_B$. For configurations with $q < 1$ we find that the total mass approaches cero as $f_0 \to 0$, then increases until it reaches a maximum, and then decreases again. However, the case $q = 1$ behaves quite differently, with the mass becoming a monotonically decreasing function of $f_0$. Moreover, perhaps surprisingly, our results seem to indicate that for $q = 1$ the total mass in fact increases without bound as $f_0 \to 0$. We have also found that in all cases with $q \leq 1$ the maximum total mass coincides with the minimum binding energy. Also, the variation of $M_{RN}$ with respect to $\omega$ describes spiral pattern for $q < 1$, this behavior is very similar to that found for the uncharged case [6–8], and for self-gravitating fields with spin $s = 0, 1$ [5, 16].

As has also been found to be the case for both boson and Proca stars [5, 16], we are able to find super-critical solutions for which $q > m = 1$, but only for a limited range of values of $f_0$ and $q$. Interestingly, we find that all our configurations are such that $Q/M_{RN} < 1$, even for those cases with $q > m = 1$. Therefore, the total charge is always smaller than the total mass, which explains why stationary solutions can still exist even for $q > m = 1$.

For all values of the charges, there exist Dirac stars with compactness of the same order of magnitude as those of neutron stars. However, gravitationally bound configurations with such compactness are found only for ($q < 1$). These results indicate the existence of highly compact gravitationally bound charged Dirac stars.

Finally, our results suggest that, at least at the purely classical level, electrically charged fields with different spins

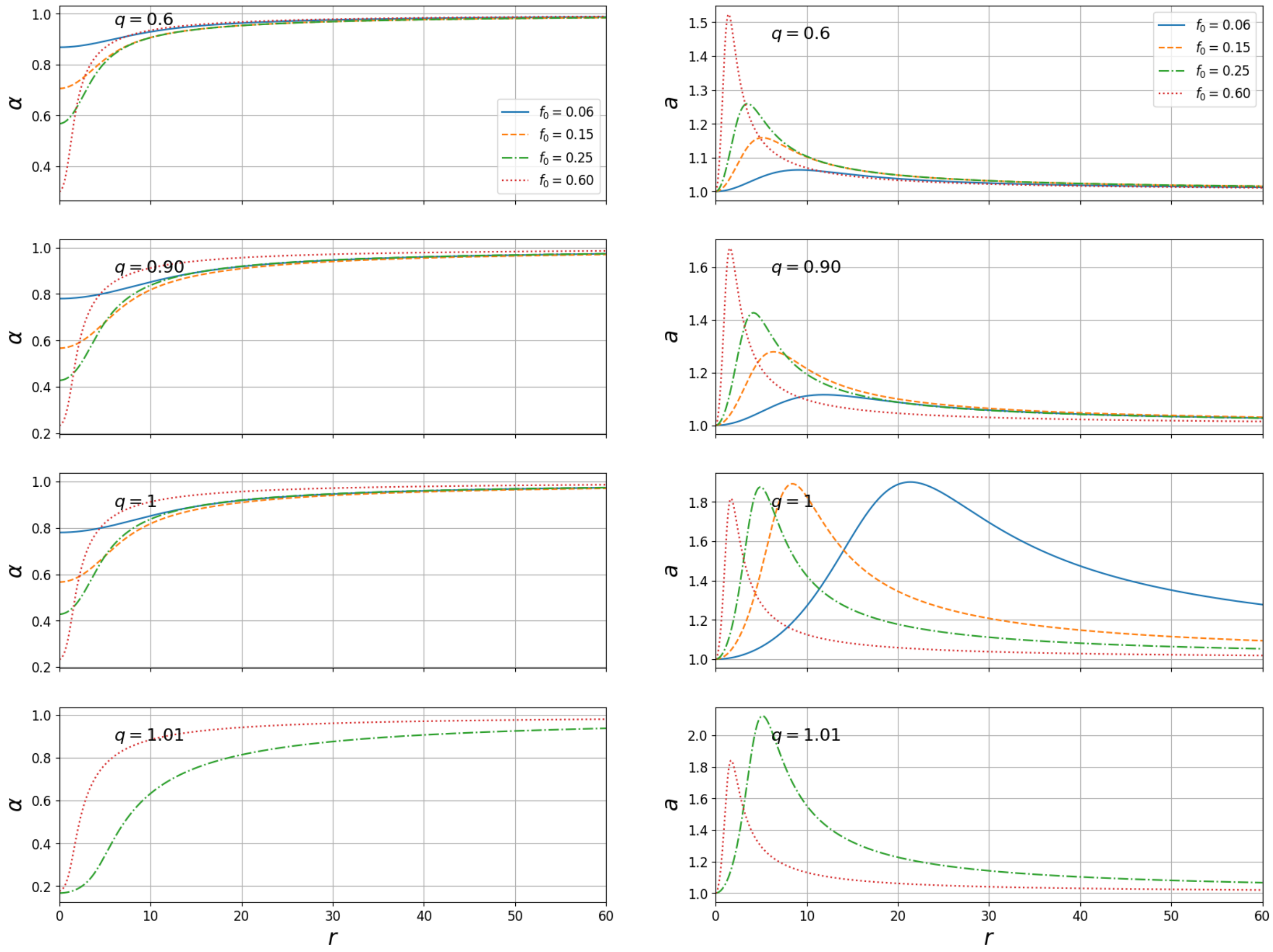

**Figure 9:** Left panel: Lapse function $\alpha(r)$. Right panel: Radial metric function $a(r)$.

$s = 0, 1/2, 1$ share a wide range of common characteristics when they coupled to gravity.

## ACKNOWLEDGMENTS

The authors wish to thank Carlos Joaquin, Yahir Mio and Elishah Candanosa for many useful discussions and comments. This work was partially supported by DGAPA-UNAM project IN100523. MH-M also acknowledges financial support from a SECIHTI postdoctoral fellowship.

---

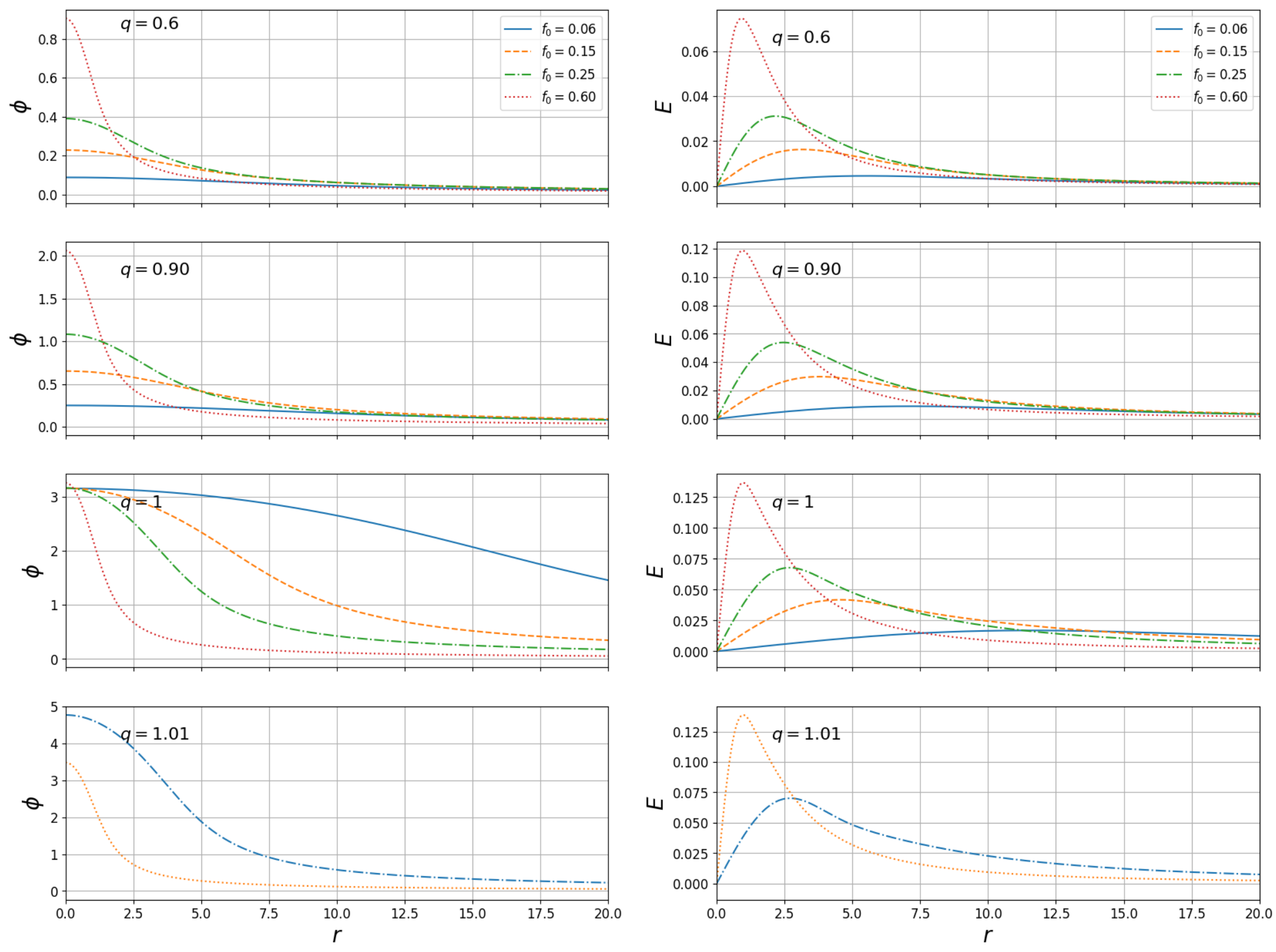

**Figure 10:** Left panel: Electric potential $\phi(r)$. Right panel: Electric field $E(r)$.

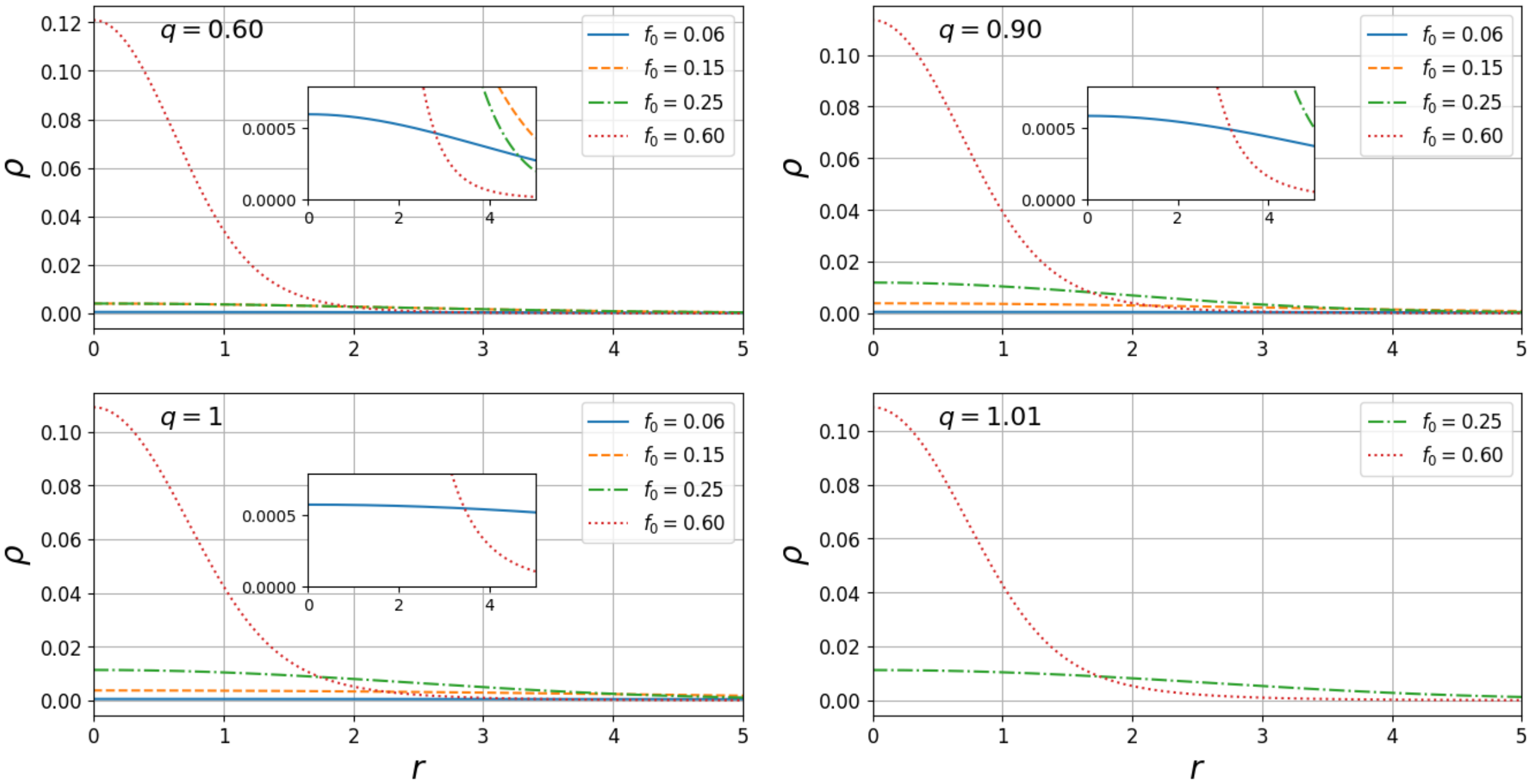

**Figure 11:** Total energy density $\rho(r)$ for four values of $q = (0.60, 0.90, 1.00, 1.01)$, and four values of the central amplitude of the Dirac field $f_0 = (0.06, 0.15, 0.25, 0.60)$. For $f_0 = 0.06$ the density is very small but is different from zero. In all cases the energy density is concentrated near the origin.